\pdfoutput=1
\documentclass[aps,prb,preprint,amsmath,amssymb,longbibliography,floatfix]{revtex4-2}

\usepackage[T1]{fontenc}
\usepackage{graphicx}
\usepackage{bm}
\usepackage{tikz}
\usetikzlibrary{arrows.meta,decorations.pathreplacing}
\usepackage[breaklinks=true,colorlinks=true,linkcolor=blue,urlcolor=blue,citecolor=blue]{hyperref}

\newcommand{\figfile}[2]{\includegraphics[width=#2]{#1}}

\newcommand{\Jt}{J_2}
\newcommand{\Jc}{J_{2c}}
\newcommand{\JL}{J_{2}^{\mathrm{L}}}
\newcommand{\Jd}{J_{2}^{\mathrm{d}}}
\newcommand{\Sv}{\mathbf{S}}
\newcommand{\Tr}{\operatorname{Tr}}
\newcommand{\Icft}{I_{\mathrm{CFT}}}
\newcommand{\xichi}{\xi_\chi}

\begin{document}

\preprint{}

\title{Mutual information across the commensurate--incommensurate crossover and near the
Kosterlitz--Thouless point of the spin-$\tfrac12$ $J_1$--$J_2$ Heisenberg chain}

\author{Wuttichai Pankeaw}
\email{pankeaw.wutti@gmail.com}
\author{Pruet Kalasuwan}
\email{pruet.k@psu.ac.th}
\affiliation{Division of Physical Science, Faculty of Science, Prince of Songkla University,
Hat Yai, Songkhla 90110, Thailand}

\begin{abstract}
Quantum mutual information captures all correlations between two subsystems, but what it adds 
to conventional correlation functions is often unclear. We address this question for the 
spin-$\tfrac12$ $J_1$--$J_2$ Heisenberg chain, using density-matrix renormalization group 
calculations on infinite chains and periodic rings together with an SU(2) decomposition of 
the mutual information into correlation channels. For SU(2)-singlet ground states the 
two-site mutual information is an exact function of the spin correlator, so any additional 
content must come from blocks. In the dimerized phase, blocks resolve bond-energy 
correlations in the SU(2)-scalar channel without introducing an independent asymptotic 
correlation length: the leading singlet and triplet correlation lengths agree to about $1\%$. 
In the commensurate regime up to $J_2/J_1=0.40$, the scalar channel accounts for most of the 
block mutual information at the largest distances computed, whereas its share falls to a few 
percent beyond the disorder point. The disorder point coincides, within numerical resolution, 
with the Majumdar--Ghosh point $J_2/J_1=\tfrac12$, while the structure-factor peak splits 
only at the Lifshitz point $J_2/J_1=0.520622(2)$. On the gapless side we compare block mutual 
information on periodic rings, extrapolated to infinite size, with the two-interval 
prediction of SU(2)$_1$ conformal field theory across the Kosterlitz--Thouless point 
$J_2/J_1=0.241167$. At this point, for blocks of $l\le6$ sites, the correction decays as 
$l^{-p}$ with $p=1.96$--$2.07$, and any logarithmic term is at least $60$ times smaller than 
in the Heisenberg chain. Expanding in the marginal coupling, measured independently by level 
spectroscopy, we find a first-order coefficient consistent with $l^{-2}$ scaling and no 
detectable $l$-independent contribution, together with a second-order coefficient that is 
positive for $x\le1/4$ over the accessible block lengths and consistent with a positive 
asymptotic limit. This suggests that the mutual information relative to the field-theory 
value becomes stationary at the Kosterlitz--Thouless point in the large-block limit. The 
share of the scalar channel approaches $1/4$ at small cross ratio $x$, as expected from the 
SO(4) symmetry of the fixed point. Under two stated assumptions, the splitting of the 
SU(2)$_1$ multiplet by the marginal coupling yields the small-$x$ prediction 
$\tfrac38\ln x-\tfrac58$ for the first-order response of this share to the coupling; for 
$l=4$, the numerical results agree with this prediction within $1$--$3\%$ at $x\le1/9$.
\end{abstract}

\maketitle

\section{Introduction}
\label{sec:intro}

Frustrated quantum magnets show how competing interactions destroy conventional order and stabilize phases without classical
counterparts~\cite{Amico2008}. In one dimension the spin-$\tfrac12$ Heisenberg chain with antiferromagnetic exchange $J_1$ between nearest
neighbors and $\Jt$ between next-nearest neighbors is the paradigmatic example. As the single ratio $\Jt/J_1$ increases, the chain passes from a
critical Luttinger liquid through a Kosterlitz--Thouless (KT) transition into a spontaneously dimerized phase with gapped spin-$\tfrac12$
excitations, and its short-range spin correlations change from commensurate to incommensurate~\cite{MajumdarGhosh1969a,Haldane1982,Affleck1989,WhiteAffleck1996}.
Because the phase diagram is known with high precision, the model is a natural benchmark for quantum-information measures of correlations.
Such measures are increasingly used to characterize quantum phases, but their reliability can only be judged where an independent
reference exists~\cite{Amico2008,Eisert2010}. Among them, the quantum mutual information between two regions is particularly informative. It
quantifies the total quantum and classical correlations between the regions~\cite{Groisman2005}, it bounds every connected correlation function
between them~\cite{Wolf2008}, and in a critical system it depends on the full operator content of the underlying conformal field
theory (CFT)~\cite{Furukawa2009,CCT2009}.

The established picture of the $J_1$--$J_2$ chain rests on three groups of results. Exact solutions and field theory fixed its landmarks. At
$\Jt=J_1/2$ the ground state is an exact product of nearest-neighbor singlets~\cite{MajumdarGhosh1969a,MajumdarGhosh1969b} whose excitations form a
two-spinon continuum~\cite{ShastrySutherland1981}. The critical phase is described by the SU(2)$_1$ Wess--Zumino--Witten theory with a marginally
irrelevant current--current coupling that produces logarithmic corrections~\cite{Haldane1982,Affleck1989}. This coupling vanishes at the KT point,
located by level spectroscopy at $\Jc/J_1=0.241167$, where the logarithmic corrections disappear~\cite{OkamotoNomura1992,Eggert1996}. Numerical
studies, mainly density-matrix renormalization group (DMRG) and exact diagonalization (ED) of finite chains, then characterized the dimerized phase:
its order parameter, gap and correlation length, and the incommensurate wave vector of spin correlations obtained from real-space correlators,
structure-factor peaks and level crossings~\cite{WhiteAffleck1996,Chitra1995,Bursill1995,Kumar2015,Soos2016}. These studies showed that the
commensurate--incommensurate crossover proceeds in two steps, a sequence known from frustrated Ising and spin-1
chains~\cite{Stephenson1970,Schollwoeck1996,Chepiga2016}. At a disorder point, which for the spin-$\tfrac12$ chain coincides with the
Majumdar--Ghosh point, the decay of correlations becomes oscillatory. At a Lifshitz point, located by infinite-system DMRG at $\Jt/J_1=0.52063(6)$~\cite{Bursill1995,RothSchollwoeck1998},
the maximum of the structure factor leaves $k=\pi$. Finally, entanglement studies established how critical chains are probed by entropies.
Single-block entropies yield the central charge~\cite{Holzhey1994,Vidal2003,CalabreseCardy2004,CalabreseCardy2009}, and the mutual information of two
disjoint intervals has been computed in closed form for the compactified boson~\cite{CCT2009,CCT2011} and tested on XXZ chains up to the
isotropic point~\cite{Furukawa2009,Alba2011}. For the $J_1$--$J_2$ chain itself, mutual information has been studied in its basis-dependent
Shannon form, which shows universal scaling at criticality~\cite{Alcaraz2013,Alcaraz2015}.

Two questions nevertheless remain open. First, the commensurate--incommensurate crossover has been characterized through the spin correlation
function and quantities derived from it, in finite chains. Whether correlation measures that also contain multi-spin correlations, such as the
mutual information between blocks of sites, display the crossover in the same way, with the same correlation length, the same wave vector and the
same onset, or reveal structure the spin correlator does not show, remains unresolved. Second, comparisons of lattice mutual information with the two-interval CFT
prediction at the SU(2) point have been carried out for the Heisenberg chain. There, slow logarithmic corrections from the marginal coupling
coexist with lattice corrections that alternate with the parity of the block length~\cite{Laflorencie2006,Calabrese2010parity,CardyCalabrese2010},
so the two cannot be separated. The $J_1$--$J_2$ chain allows the marginal coupling to be tuned through zero, and for the single-block entropy
this has been used to remove the logarithmic part of the alternating correction~\cite{Laflorencie2006}. For the mutual information of two
intervals, however, it is not known how the approach to the CFT prediction changes as the coupling passes through $\Jc$, whether the logarithmic
corrections indeed vanish there, or how the correction depends on the marginal coupling on either side of $\Jc$.

Both questions call for calculations directly in the thermodynamic limit, with access to correlation lengths, wave vectors and multi-site reduced
density matrices under controlled conditions. Infinite-system DMRG (iDMRG)~\cite{White1992,McCulloch2008,Schollwoeck2011} provides
translation-invariant matrix-product states free of boundaries and finite-size effects. The transfer matrix of such a state yields correlation
lengths and incommensurate wave vectors without fitting correlation functions, and it can separate spin correlations from dimer correlations. The
finite-bond-dimension bias of these quantities is understood and can be extrapolated~\cite{Tagliacozzo2008,Pollmann2009,Rams2018}. Because the
ground state is an SU(2) singlet, the correlated part of a two-block density matrix can moreover be decomposed exactly into spin, scalar and
higher-spin channels, where the scalar channel contains the bond-energy (dimer) correlations, which makes it possible to state precisely which
correlations the mutual information contains. On the critical side,
comparing the mutual information with the CFT prediction at a fixed four-point cross ratio while the block size grows isolates the corrections, and
the marginal coupling responsible for them can be measured independently by level spectroscopy on finite rings~\cite{OkamotoNomura1992}. Periodic
rings, extrapolated to infinite size, are free of the spurious dimerization that finite-bond-dimension infinite states develop near criticality, and
they can be followed through $\Jc$ into the weakly dimerized phase, where the coupling changes sign. Data on both sides of $\Jc$ separate the orders
of the expansion in the coupling.

In this work, we use iDMRG, DMRG on periodic rings, transfer-matrix analysis, exact diagonalization and an SU(2) channel decomposition to determine what mutual information
adds to conventional spin correlations in the spin-$\tfrac12$ $J_1$--$J_2$ chain. We examine the commensurate--incommensurate crossover of the dimerized
phase, including the disorder and Lifshitz points, and the approach of block mutual information to the two-interval CFT prediction on both sides of the
KT point. This combination allows us to separate correlation-length information from channel content, and to test the effect of the marginal coupling
directly, with an independently measured coupling and a controlled estimate of finite-bond-dimension effects.

The paper is organized as follows. Section~\ref{sec:theory} introduces the model, the mutual-information quantities, the CFT predictions and the
analytical tools. Section~\ref{sec:methods} describes the numerical methods. The results are presented and discussed in Sec.~\ref{sec:results},
and Sec.~\ref{sec:conclusions} summarizes our conclusions. Technical details are collected in the Appendices.

\section{Model and theory}
\label{sec:theory}

\subsection{Hamiltonian and phase diagram}
\label{sec:model}

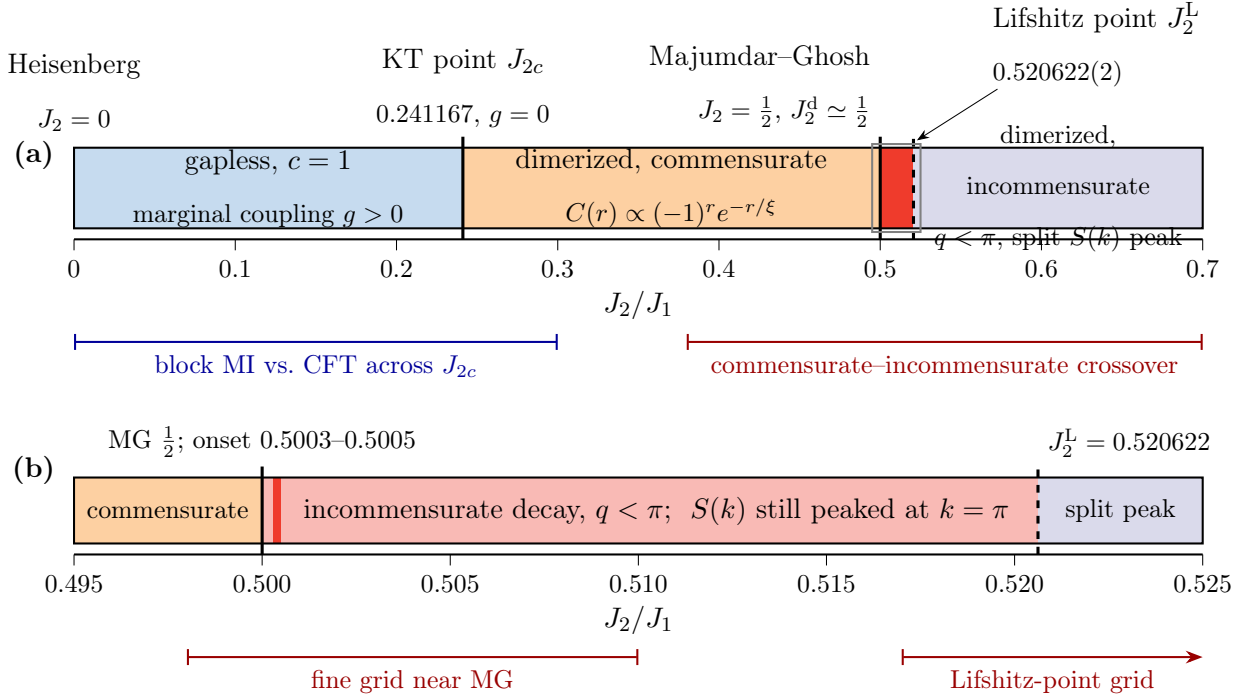
\begin{figure*}[t]
\centering
\resizebox{\textwidth}{!}{%
\begin{tikzpicture}[>=Stealth, font=\small]
  \definecolor{cgapless}{RGB}{198,219,239}
  \definecolor{ccomm}{RGB}{253,208,162}
  \definecolor{cwindow}{RGB}{239,59,44}
  \definecolor{cinc}{RGB}{218,218,235}
  \fill[cgapless] (0,0) rectangle (5.3057,1.1);
  \fill[ccomm]    (5.3057,0) rectangle (11.0,1.1);
  \fill[cwindow]  (11.0,0) rectangle (11.4537,1.1);
  \fill[cinc]     (11.4537,0) rectangle (15.4,1.1);
  \draw[thick] (0,0) rectangle (15.4,1.1);
  \node[align=center] at (2.65,0.55) {gapless, $c=1$\\[-1pt]\footnotesize marginal coupling $g>0$};
  \node[align=center] at (8.15,0.55) {dimerized, commensurate\\[-1pt]\footnotesize $C(r)\propto(-1)^r e^{-r/\xi}$};
  \node[align=center] at (13.43,0.55) {\footnotesize dimerized,\\[-2pt]\footnotesize incommensurate\\[-2pt]\footnotesize $q<\pi$, split $S(k)$ peak};
  \draw[very thick] (5.3057,-0.15) -- (5.3057,1.25);
  \draw[very thick] (11.0,-0.15) -- (11.0,1.25);
  \draw[very thick, dashed] (11.4537,-0.15) -- (11.4537,1.25);
  \node[anchor=south, align=center] at (0,1.2) {Heisenberg\\[-1pt]\footnotesize $\Jt=0$};
  \node[anchor=south, align=center] at (5.3057,1.25) {KT point $\Jc$\\[-1pt]\footnotesize $0.241167$, $g=0$};
  \node[anchor=south east, align=right] at (11.0,1.25) {Majumdar--Ghosh\\[-1pt]\footnotesize $\Jt=\tfrac12$, $\Jd\simeq\tfrac12$};
  \draw[->] (12.6,1.9) -- (11.5,1.2);
  \node[anchor=south west, align=left] at (12.4,1.8) {Lifshitz point $\JL$\\[-1pt]\footnotesize $0.520622(2)$};
  \draw[thick] (0,-0.15) -- (15.4,-0.15);
  \foreach \x/\l in {0/0, 2.2/0.1, 4.4/0.2, 6.6/0.3, 8.8/0.4, 11.0/0.5, 13.2/0.6, 15.4/0.7}
    {\draw (\x,-0.15) -- (\x,-0.3); \node[anchor=north] at (\x,-0.3) {\footnotesize $\l$};}
  \node[anchor=north] at (7.7,-0.72) {$\Jt/J_1$};
  \node[font=\bfseries] at (-0.55,1.0) {(a)};
  \draw[|-|, thick, blue!60!black] (0,-1.55) -- (6.6,-1.55);
  \node[anchor=north, blue!60!black] at (3.3,-1.6) {\footnotesize block MI vs.\ CFT across $\Jc$};
  \draw[|-|, thick, red!60!black] (8.36,-1.55) -- (15.4,-1.55);
  \node[anchor=north, red!60!black] at (11.88,-1.6) {\footnotesize commensurate--incommensurate crossover};
  \draw[gray, thick] (10.89,-0.05) rectangle (11.55,1.15);
  \begin{scope}[yshift=-4.3cm]
    \fill[ccomm]   (0,0) rectangle (2.5667,0.9);
    \fill[cwindow!35] (2.5667,0) rectangle (13.1526,0.9);
    \fill[cinc]    (13.1526,0) rectangle (15.4,0.9);
    \fill[cwindow] (2.7207,0) rectangle (2.8233,0.9);
    \draw[thick] (0,0) rectangle (15.4,0.9);
    \node[align=center] at (1.28,0.45) {\footnotesize commensurate};
    \node[align=center] at (7.95,0.45) {incommensurate decay, $q<\pi$; \ $S(k)$ still peaked at $k=\pi$};
    \node[align=center] at (14.28,0.45) {\footnotesize split peak};
    \draw[very thick] (2.5667,-0.15) -- (2.5667,1.05);
    \node[anchor=south, align=center] at (2.5667,1.05) {\footnotesize MG $\tfrac12$; onset $0.5003$--$0.5005$};
    \draw[very thick, dashed] (13.1526,-0.15) -- (13.1526,1.05);
    \node[anchor=south west, align=left] at (13.1526,1.05) {\footnotesize $\JL=0.520622$};
    \draw[thick] (0,-0.15) -- (15.4,-0.15);
    \foreach \x/\l in {0/0.495, 2.5667/0.500, 5.1333/0.505, 7.7/0.510, 10.2667/0.515, 12.8333/0.520, 15.4/0.525}
      {\draw (\x,-0.15) -- (\x,-0.3); \node[anchor=north] at (\x,-0.3) {\footnotesize $\l$};}
    \node[anchor=north] at (7.7,-0.72) {$\Jt/J_1$};
    \node[font=\bfseries] at (-0.55,1.0) {(b)};
    \draw[|-|, thick, red!60!black] (1.54,-1.55) -- (7.7,-1.55);
    \node[anchor=north, red!60!black] at (4.62,-1.6) {\footnotesize fine grid near MG};
    \draw[|->, thick, red!60!black] (11.293,-1.55) -- (15.4,-1.55);
    \node[anchor=north, red!60!black] at (13.35,-1.6) {\footnotesize Lifshitz-point grid};
  \end{scope}
\end{tikzpicture}}
\caption{Phase diagram of the spin-$\tfrac12$ $J_1$--$J_2$ chain for $0\le\Jt/J_1\le0.7$. (a) Gapless phase with a marginally
irrelevant coupling $g>0$ for $\Jt<\Jc$; Kosterlitz--Thouless transition at $\Jc=0.241167$~\cite{OkamotoNomura1992,Eggert1996}; dimerized phase with
commensurate (orange) and incommensurate (lavender) short-range order, separated by the narrow window (red) between the disorder point
$\Jd\simeq\tfrac12$, where the real-space decay becomes incommensurate, and the Lifshitz point $\JL$, where the peak of $S(k)$ leaves $k=\pi$. (b) Magnification of the gray box in (a), $0.495\le\Jt\le0.525$. Dark red bar: onset of incommensurate decay
found here ($0.5003<\Jd\le0.5005$); dashed line: Lifshitz point.
Bars below the axes indicate the parameter ranges studied in this work (MI: mutual information; CFT: conformal field theory).}
\label{fig:phasediagram}
\end{figure*}

We consider the spin-$\tfrac12$ Heisenberg chain with antiferromagnetic exchange between nearest and next-nearest neighbors,
\begin{equation}
H = J_1\sum_i \Sv_i\cdot\Sv_{i+1} + \Jt\sum_i \Sv_i\cdot\Sv_{i+2},
\label{eq:H_intro}
\end{equation}
on an infinite chain, and set $J_1=1$ so that $\Jt$ denotes the ratio $\Jt/J_1$. The Hamiltonian commutes with the total spin. The
ground state is therefore an SU(2) singlet throughout the parameter range studied here; in the dimerized phase this holds for each of the
two symmetry-broken ground states separately. The phase diagram for $0\le\Jt\le0.7$ is shown in Fig.~\ref{fig:phasediagram}.

For $\Jt<\Jc$ the chain is critical. Its low-energy theory is the SU(2)$_1$ Wess--Zumino--Witten model, equivalently a compactified boson
with central charge $c=1$ at the SU(2)-symmetric radius~\cite{Haldane1982,Affleck1989}. The leading perturbation is a marginal
current--current interaction with coupling $g$, which vanishes at $\Jc$. For $\Jt<\Jc$ the coupling is
marginally irrelevant and flows to zero only logarithmically, which produces multiplicative logarithmic corrections to all critical
quantities~\cite{Affleck1989,Eggert1996}. At $\Jc=0.241167$, determined by level spectroscopy~\cite{OkamotoNomura1992}, the coupling vanishes
and these corrections disappear~\cite{Eggert1996}. For $\Jt>\Jc$ the coupling is marginally relevant: a spin gap opens through a
Kosterlitz--Thouless (KT) transition, exponentially small close to $\Jc$, and translation symmetry is spontaneously broken by dimerization.
The elementary excitations of the dimerized phase are spin-$\tfrac12$ domain walls between the two dimer patterns (spinons), so that
excitations with total spin $0$ and $1$ form two-spinon continua~\cite{ShastrySutherland1981}.

At the Majumdar--Ghosh (MG) point $\Jt=\tfrac12$ the two ground states are exact products of nearest-neighbor singlets and all spin
correlations beyond one bond vanish~\cite{MajumdarGhosh1969a,MajumdarGhosh1969b}. This point also organizes the short-range order of the
dimerized phase into three regimes~\cite{WhiteAffleck1996,Bursill1995,Soos2016}. For $\Jc<\Jt<\Jd$ correlations decay as a staggered
exponential, $C(r)\propto(-1)^re^{-r/\xi}$, and the static structure factor $S(k)$ is peaked at $k=\pi$. At the disorder point
$\Jd$~\cite{Stephenson1970,Schollwoeck1996} the asymptotic decay becomes oscillatory, $C(r)\propto\cos(qr+\varphi)e^{-r/\xi}$ with $q<\pi$,
while $S(k)$ still has its maximum at $\pi$. At the Lifshitz point $\JL$ the maximum of $S(k)$ splits and moves away from $\pi$; beyond it
both real-space decay and structure factor are incommensurate, and $q$ decreases continuously towards $\pi/2$ as $\Jt$ grows. In this work we
find $\Jd$ between $0.5003$ and $0.5005$, i.e.\ at the MG point within our resolution, and $\JL=0.520622(2)$, in agreement with the infinite-system DMRG value $0.52063(6)$ of Ref.~\cite{Bursill1995}
(Sec.~\ref{sec:Lifshitz}). The window $\Jd<\Jt<\JL$, magnified in
Fig.~\ref{fig:phasediagram}(b), is the regime in which real-space and momentum-space probes disagree on whether the state is commensurate.

The calculations of this work cover two ranges, marked below the axis of Fig.~\ref{fig:phasediagram}: the gapless side and the vicinity of the KT
point, $0\le\Jt\le0.30$, where block mutual information is compared with the field-theory prediction on both sides of $\Jc$, and the dimerized phase $0.38\le\Jt\le0.70$, where the
commensurate--incommensurate crossover is studied, supplemented by fine grids around the MG and Lifshitz points.

\subsection{Mutual information}
\label{sec:MI}

\begin{figure}[!htb]
\centering
\resizebox{\ifdim\columnwidth>12cm 0.65\columnwidth\else\columnwidth\fi}{!}{%
\begin{tikzpicture}[font=\large,
    site/.style={circle, draw=black, fill=white, inner sep=0pt, minimum size=6pt},
    siteA/.style={circle, draw=black, fill=red!70!black, inner sep=0pt, minimum size=8pt},
    siteB1/.style={circle, draw=black, fill=blue!70!black, inner sep=0pt, minimum size=8pt},
    siteB2/.style={circle, draw=black, fill=orange!90!black, inner sep=0pt, minimum size=8pt}]
  \def\a{0.8}
  \node[font=\Large\bfseries] at (-1.3,0) {(i)};
  \foreach \k in {0,...,11} {\draw[gray!55] (\k*\a,0) to[bend right=55] ({(\k+2)*\a},0);}
  \draw[gray!80!black] (0,0) -- (13*\a,0);
  \draw[gray!80!black, dotted] (-0.6,0) -- (0,0); \draw[gray!80!black, dotted] (13*\a,0) -- (13*\a+0.6,0);
  \foreach \k in {0,...,13} {\node[site] at (\k*\a,0) {};}
  \node[siteA] at (3*\a,0) {};
  \node[siteA] at (10*\a,0) {};
  \node[anchor=north] at (3*\a,-0.62) {$i$};
  \node[anchor=north] at (10*\a,-0.62) {$j=i+r$};
  \draw[decorate, decoration={brace, amplitude=6pt, raise=5pt}] (3*\a,0) -- (10*\a,0) node[midway, above=11pt] {$r$};
  \draw[gray!80!black] (-0.5,1.35) -- (0.3,1.35); \node[anchor=west] at (0.35,1.35) {$J_1$};
  \draw[gray!55] (1.4,1.25) to[bend left=40] (2.2,1.25); \node[anchor=west] at (2.25,1.35) {$J_2$};
  \node[anchor=east] at (13*\a+0.6,1.35) {$I(i{:}j)=S_i+S_j-S_{ij}$};
  \begin{scope}[yshift=-3.4cm]
    \node[font=\Large\bfseries] at (-1.3,0) {(ii)};
    \fill[blue!12, rounded corners=3pt]   (1.6*\a,-0.32) rectangle (4.4*\a,0.32);
    \fill[orange!18, rounded corners=3pt] (6.6*\a,-0.32) rectangle (9.4*\a,0.32);
    \draw[blue!70!black, rounded corners=3pt]   (1.6*\a,-0.32) rectangle (4.4*\a,0.32);
    \draw[orange!90!black, rounded corners=3pt] (6.6*\a,-0.32) rectangle (9.4*\a,0.32);
    \foreach \k in {0,...,12} {\draw[gray!80!black] (\k*\a,0) -- ({(\k+1)*\a},0);}
    \foreach \k in {0,2,...,12} {\draw[line width=2.4pt, gray!65!black] (\k*\a,0) -- ({(\k+1)*\a},0);}
    \draw[gray!80!black, dotted] (-0.6,0) -- (0,0); \draw[gray!80!black, dotted] (13*\a,0) -- (13*\a+0.6,0);
    \foreach \k in {0,...,13} {\node[site] at (\k*\a,0) {};}
    \foreach \k in {2,3,4} {\node[siteB1] at (\k*\a,0) {};}
    \foreach \k in {7,8,9} {\node[siteB2] at (\k*\a,0) {};}
    \draw[decorate, decoration={brace, amplitude=6pt, raise=10pt}] (1.8*\a,0) -- (4.2*\a,0) node[midway, above=16pt] {$l$};
    \draw[decorate, decoration={brace, amplitude=6pt, raise=10pt}] (4.8*\a,0) -- (6.2*\a,0) node[midway, above=16pt] {$d$};
    \draw[decorate, decoration={brace, amplitude=6pt, raise=10pt}] (6.8*\a,0) -- (9.2*\a,0) node[midway, above=16pt] {$l$};
    \node[anchor=north, blue!70!black]   at (3*\a,-0.45) {$A_1$};
    \node[anchor=north, orange!90!black] at (8*\a,-0.45) {$A_2$};
    \draw[->, >=stealth, thick] (2*\a,-1.35) -- (2*\a,-0.4);
    \node[anchor=north] at (2*\a,-1.3) {$s_0$};
    \node[anchor=north, gray!50!black] at (11.5*\a,-0.45) {environment};
    \draw[line width=2.4pt, gray!65!black] (-0.5,1.45) -- (0.3,1.45); \node[anchor=west] at (0.35,1.45) {strong bond};
    \node[anchor=east] at (13*\a+0.6,1.45) {$I_l(d)=S_{A_1}+S_{A_2}-S_{A_1\cup A_2}$};
  \end{scope}
\end{tikzpicture}}
\caption{Geometries used for the mutual information. (i) Two sites $i$ and $j=i+r$ of the $J_1$--$J_2$ chain; straight lines are
nearest-neighbor ($J_1$) bonds and arcs next-nearest-neighbor ($J_2$) bonds. (ii) Two blocks $A_1$ and $A_2$ of $l$ sites each, separated by
$d$ sites, shown for $l=3$ and $d=2$; the rest of the infinite chain (environment) is traced out. Thick lines indicate the strong bonds of one
dimerization pattern. In the dimerized phase the result depends on whether the first site $s_0$ of $A_1$ is the left site of a strong bond
($s_0$ even, as shown) or of a weak bond ($s_0$ odd).}
\label{fig:geometry}
\end{figure}
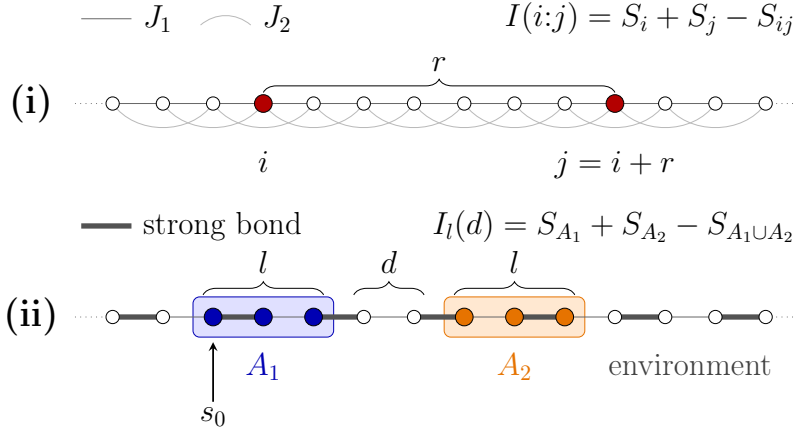

For a region $X$ with reduced density matrix $\rho_X=\Tr_{\bar X}|\psi\rangle\langle\psi|$ we use the von Neumann entropy
$S_X=-\Tr\rho_X\ln\rho_X$ and the Rényi-2 entropy $S^{(2)}_X=-\ln\Tr\rho_X^2$. We consider the two geometries shown in
Fig.~\ref{fig:geometry}: (i) two sites $i$ and $j=i+r$, with $I(i{:}j)=S_i+S_j-S_{ij}$; (ii) two blocks of equal length $l$,
$A_1=[s_0,s_0+l)$ and $A_2=[s_0+l+d,\,s_0+2l+d)$, separated by $d$ sites, with
\begin{equation}
I_l(d) = S_{A_1}+S_{A_2}-S_{A_1\cup A_2}.
\label{eq:blockMI}
\end{equation}
In the dimerized phase translation symmetry is broken down to translations by two sites, so we evaluate both starting sublattices
$s_0\in\{0,1\}$ and report either both or their average.

\paragraph*{Two-site mutual information of SU(2) singlets.}
SU(2) invariance restricts the two-site reduced density matrix of a singlet to a mixture of the singlet projector $P_s$ and the triplet
projector $P_t$, i.e.\ a Werner state~\cite{Werner1989}:
\begin{equation}
\rho_{ij}=\lambda_s P_s+\lambda_t P_t,\qquad \lambda_s=\tfrac14-C,\quad \lambda_t=\tfrac14+\tfrac{C}{3},
\label{eq:werner}
\end{equation}
where $C\equiv\langle\Sv_i\cdot\Sv_j\rangle$ and $\lambda_t$ is threefold degenerate. Each single-site density matrix is $\mathbb{1}/2$.
Hence
\begin{equation}
I(i{:}j)=f(C)\equiv 2\ln2+\lambda_s\ln\lambda_s+3\lambda_t\ln\lambda_t\simeq\frac{8}{3}C^2+O(C^3)
\label{eq:fC}
\end{equation}
(derivation in Appendix~\ref{app:fC}). Two consequences follow. First, two-site mutual information contains no information beyond the spin
correlator; any feature it shows is a feature of $C(r)$. Second, since $f\propto C^2$ for weak correlations, an incommensurate decay
$C(r)\propto\cos(qr+\varphi)e^{-r/\xi}$ produces a mutual information $\propto\cos^2(qr+\varphi)e^{-2r/\xi}$, i.e.\ a modulation with wave vector
$2q$ and true nodes. Neither statement holds for blocks with $l\ge2$, whose reduced density matrices also contain multi-spin (e.g.\
dimer--dimer) correlations.

\subsection{SU(2) channel decomposition of block mutual information}
\label{sec:channels}

To identify which correlations block mutual information contains, we decompose the correlated part of the two-block density matrix,
\begin{equation}
G\equiv\rho_{A_1A_2}-\rho_{A_1}\otimes\rho_{A_2},
\label{eq:G}
\end{equation}
into irreducible SU(2) channels. The operator $G$ is the correlation density matrix introduced by Cheong and Henley~\cite{CheongHenley2009},
whose singular-value structure has been analyzed with DMRG for one-dimensional chains, including its symmetry sectors~\cite{Munder2010}. The new
elements here are a weighting that turns the channel content into additive shares of the mutual information, and its systematic dependence on $\Jt$
and on the block geometry. Operators acting on a block of $l$ spins decompose into irreducible tensor operators of rank $k=0,1,\dots,l$, which
are eigenvectors of the adjoint Casimir superoperator $\mathcal{C}(O)=\sum_a[S^a_A,[S^a_A,O]]$ with eigenvalue $k(k+1)$, where $\Sv_A$ is the total
spin of the block. Rank $1$ contains the spin operators and rank $k\ge2$ multi-spin operators. Rank $0$ is the SU(2)-scalar sector. For the block sizes
used here it is spanned by explicitly known operators: for $l=2$ by the identity and the bond energy $\Sv_1\cdot\Sv_2$ only, and for $l=3$ by the
identity, the three bond energies $\Sv_1\cdot\Sv_2$, $\Sv_2\cdot\Sv_3$, $\Sv_1\cdot\Sv_3$ and the scalar chirality $\Sv_1\cdot(\Sv_2\times\Sv_3)$
(Appendix~\ref{app:channeldecomp}). For $l=2$ the scalar channel therefore contains bond-energy correlations only; for $l=3$ the chirality, which is
odd under time reversal, cannot correlate with the bond energies, and its contribution is computed separately. For a singlet, $G$ is invariant under
global rotations, so only equal ranks on $A_1$ and $A_2$ are paired, $G=\sum_k G_k$ with $G_k=(\mathcal{P}_k\otimes\mathcal{P}_k)G$ and $\mathcal{P}_k$
the projector onto rank $k$. For $l=1$ only $k=1$ exists, in agreement with Eq.~\eqref{eq:fC}.

To assign a share of the mutual information to each channel we use $I(A_1{:}A_2)=S(\rho_{A_1A_2}\Vert\tau)$ with $\tau=\rho_{A_1}\otimes\rho_{A_2}$.
Expanding the relative entropy to second order in $G$ gives
\begin{equation}
I\simeq\tfrac12\langle G,\Omega_\tau(G)\rangle=\sum_k I_k,\qquad
I_k=\frac12\sum_{ij}\bigl|(G_k)_{ij}\bigr|^2\,\frac{\ln\tau_i-\ln\tau_j}{\tau_i-\tau_j},
\label{eq:channelMI}
\end{equation}
where $\Omega_\tau$ is the Kubo--Mori metric, written in the eigenbasis of $\tau$ with eigenvalues $\tau_i$. Because $\tau$ is invariant under
independent rotations of $A_1$ and $A_2$, channels of different rank are orthogonal in this metric and the sum is exact at second order. Within the scalar
channel the bond-energy and chirality parts are not orthogonal in this metric; we report their individual contributions and the remaining interference
term. The expansion holds when correlations are weak, i.e.\ at large $d$; we use only distances where $\sum_kI_k$ agrees with the exact $I$ within
$10\%$, report the shares $f_k=I_k/\sum_{k'}I_{k'}$, and estimate their systematic uncertainty by varying this acceptance threshold between $5\%$ and
$15\%$ and the range of distances used.

\subsection{Conformal-field-theory prediction for two intervals}
\label{sec:CFT}

For two intervals of length $l$ separated by $d$ on an infinite line, CFT predicts that the Rényi mutual information depends on the
four-point cross ratio
\begin{equation}
x=\frac{l^2}{(l+d)^2},
\label{eq:crossratio}
\end{equation}
through~\cite{CCT2009,CCT2011}
\begin{equation}
I^{(n)}=-\frac{c}{6}\Big(1+\frac1n\Big)\ln(1-x)+\frac{1}{n-1}\ln\mathcal{F}_n(x),
\label{eq:CFTn}
\end{equation}
where $\mathcal{F}_n$ is a model-dependent function; for the compact boson it depends on the compactification radius through a
parameter $\eta$ and is invariant under $\eta\to1/\eta$~\cite{CCT2009,CCT2011,Alba2011}. At the self-dual radius, which corresponds to the
SU(2)-symmetric Heisenberg point, $\mathcal{F}_n\equiv1$ in the normalization of Refs.~\cite{CCT2009,CCT2011}. With $c=1$ this gives the
reference values
\begin{equation}
\Icft=-\tfrac13\ln(1-x),\qquad I^{(2)}_{\mathrm{CFT}}=-\tfrac14\ln(1-x).
\label{eq:Icft}
\end{equation}
We compare the lattice data through the ratio $R=I/\Icft$ at fixed $x$, choosing $d=l$, $2l$ and $l/2$ so that $x=1/4$, $1/9$ and $4/9$ stay fixed while
$l$ grows. This removes the explicit $l$ dependence of the universal part and isolates the corrections.

\paragraph*{Corrections and the marginal coupling.}
Three types of corrections are expected. (a) Irrelevant operators of the SU(2)$_1$ theory with scaling dimension $4$ generate corrections
$\propto l^{-2}$. (b) Lattice ``unusual'' corrections, which in Rényi entropies of gapless chains appear with alternating sign and decay
with exponents set by the Luttinger parameter and the Rényi index~\cite{Laflorencie2006,Calabrese2010parity,CardyCalabrese2010};
in the von Neumann quantities used here they are found to be small. (c) The marginal current--current coupling $g$ obeys, to leading
order, $dg/d\ln L=-b\,g^2$, so that at length scale $L$
\begin{equation}
\frac{1}{g(L)}=\frac{1}{g_0}+b\ln L,
\label{eq:running}
\end{equation}
with $g_0>0$ for $\Jt<\Jc$~\cite{Affleck1989,Eggert1996}. For $\Jt<\Jc$ this produces corrections that vanish only as
$1/\ln L$; at $\Jc$ the marginal correction is absent and only (a) and (b) remain. For $\Jt>\Jc$, $g_0<0$ and $|g|$ grows with the length scale until
the gap opens; in the range used here ($\Jt\le0.28$, $|g_0|\le0.062$) the gap is far beyond the ring sizes and $|g|$ remains small.

\paragraph*{Measuring the marginal coupling.}
Instead of assuming that the coupling is proportional to the bare distance $\Jc-\Jt$, we measure it by level spectroscopy on periodic rings of $N$ sites~\cite{OkamotoNomura1992}.
To first order in $g$ the lowest triplet and the lowest excited singlet at momentum $\pi$ have scaling dimensions $\Delta_{\rm t}=\tfrac12-\tfrac g4$ and
$\Delta_{\rm s}=\tfrac12+\tfrac{3g}{4}$, so that $\Delta_{\rm s}-\Delta_{\rm t}=g$ while $3\Delta_{\rm t}+\Delta_{\rm s}=2$ is independent of $g$. With excitation
energies $E_{\rm t,s}=2\pi v\,\Delta_{\rm t,s}/N$ the
velocity drops out:
\begin{equation}
g(N)=\frac{2\,(E_s-E_t)}{E_s+3E_t}.
\label{eq:gN}
\end{equation}
The zero of $g(N)$ locates $\Jc$ for each $N$, and Eq.~\eqref{eq:running} with $L=N$ yields $g_0$ and $b$; a single marginal coupling requires the
slope $b$ to be independent of $\Jt$. We write Eq.~\eqref{eq:running} as $g_0=g(N)/[1-b\,g(N)\ln N]$, which is regular through $g=0$ and is used
unchanged on both sides of $\Jc$, and extrapolate $g_0$ linearly in $1/N^2$ (Appendix~\ref{app:coupling}).

\paragraph*{Expansion in the coupling at fixed geometry.}
At fixed block length $l$, distance $d$ and ring size, $R$ is an analytic function of $\Jt$ near $\Jc$. We therefore expand
\begin{equation}
R(\Jt;l,x)=F_0(l,x)+F_1(l,x)\,g_0+F_2(l,x)\,g_0^2+O(g_0^3),
\label{eq:Fexp}
\end{equation}
with $g_0$ the bare coupling from level spectroscopy, positive below and negative above $\Jc$; data on both sides separate odd and even orders. To first
order $g_0$ and $g(L)$ coincide, so $F_1$ does not depend on the reference scale of the coupling, whereas $F_2$ changes by a multiple of $F_1$ when that
scale is changed [Eq.~\eqref{eq:running}]. A correction of first order in the running coupling, $R-R_c=\Phi(x)\,g(l)$, would give $F_1\to\Phi(x)$ for
large $l$ and a second-order coefficient $F_2\simeq-b\,\Phi(x)\ln l$ that grows logarithmically.

\subsection{Leading small-$x$ term and the splitting of the SU(2)$_1$ multiplet}
\label{sec:smallx}

For widely separated intervals the mutual information is dominated by the primary fields of lowest scaling dimension. A scalar primary $O$ of
dimension $\Delta_O$ contributes~\cite{CCT2011,Cardy2013}
\begin{equation}
I\simeq\sum_O c(\Delta_O)\,x^{2\Delta_O},\qquad c(\Delta)=\frac{\sqrt\pi}{4}\,\frac{\Gamma(2\Delta+1)}{\Gamma(2\Delta+\tfrac32)}\,4^{-2\Delta}.
\label{eq:Cardy}
\end{equation}
At the SU(2)$_1$ point the lowest primaries are the four components of the matrix field of dimension $\tfrac12$: the three components of the staggered
magnetization and the dimerization~\cite{Affleck1989}. Since $4c(\tfrac12)=\tfrac13$, Eq.~\eqref{eq:Cardy} reproduces the leading term $x/3$ of
$\Icft$ [Eq.~\eqref{eq:Icft}]; the dimerization, the only SU(2) scalar among the four, carries one quarter of it. This is the origin of the scalar share
$1/4$ expected at small $x$.

The marginal coupling splits this multiplet into the singlet (dimerization) and the triplet (staggered magnetization) with the dimensions
$\Delta_{\rm s}=\tfrac12+\tfrac34g$ and $\Delta_{\rm t}=\tfrac12-\tfrac14g$ given above Eq.~\eqref{eq:gN}, so that $g=\Delta_{\rm s}-\Delta_{\rm t}$ at first
order in the normalization used for the measured coupling. Inserting these dimensions into Eq.~\eqref{eq:Cardy} has two consequences.
(i) The first-order shifts add up to zero over the multiplet, $\tfrac34-3\times\tfrac14=0$, so the leading term of $R$ contains no term linear in $g$.
(ii) The scalar share of the leading term, $f_0=c(\Delta_{\rm s})x^{2\Delta_{\rm s}}/[c(\Delta_{\rm s})x^{2\Delta_{\rm s}}+3c(\Delta_{\rm t})x^{2\Delta_{\rm t}}]$,
does change at first order,
\begin{equation}
\frac{\partial f_0}{\partial g}\Big|_{g=0}=\frac38\ln x+\frac{3}{16}\,\frac{d\ln c}{d\Delta}\Big|_{\Delta=\frac12}=\frac38\ln x-\frac58 ,
\label{eq:shareslope}
\end{equation}
where $d\ln c/d\Delta=2\psi(2)-2\psi(\tfrac52)-2\ln4=-\tfrac{10}{3}$ at $\Delta=\tfrac12$, with $\psi$ the digamma function. The same splitting
gives a second derivative $F_2^{\rm split}=\tfrac{3}{32}\big[(2\ln x-\tfrac{10}{3})^2+d^2\ln c/d\Delta^2\big]$ of the relative mutual information, with
$d^2\ln c/d\Delta^2=0.618$ at $\Delta=\tfrac12$, i.e.\ $3.55$, $5.66$ and $7.45$ for $x=1/4$, $1/9$ and $1/16$. Second-order shifts of the dimensions are
not included in $F_2^{\rm split}$, which therefore indicates the expected size of $F_2$ but is not a prediction.

Statements (i) and (ii) rest on assumptions that we have not proven: that at first order the marginal coupling enters the leading term only through
the scaling dimensions, i.e.\ that there are no additional $O(g)$ contributions from the branch points of the replica geometry, and that the scalar share
of the second-order expansion~\eqref{eq:channelMI} equals the share of the singlet in Eq.~\eqref{eq:Cardy}. Both refer to small $x$; at finite $x$ fields
of higher dimension contribute.

\subsection{Transfer-matrix analysis of correlations}
\label{sec:TM}

An infinite matrix-product state (iMPS) with a two-site unit cell, written in right-canonical form with tensors $B^{s}_{[1]},B^{s}_{[2]}$,
defines the unit-cell transfer map
\begin{equation}
\mathcal{T}(X)=\sum_{s_1,s_2}B^{s_1}_{[1]}B^{s_2}_{[2]}\,X\,\big(B^{s_1}_{[1]}B^{s_2}_{[2]}\big)^{\dagger}.
\label{eq:TM}
\end{equation}
Its leading eigenvalue is $1$; any two-point function decomposes as
$\langle O_0O_{2n}\rangle-\langle O\rangle^2=\sum_{k\ge2}w_k(O)\lambda_k^{\,n-1}$, where the weight
$w_k(O)=(\ell_O\cdot v_k)(u_k\cdot r_O)/(u_k\cdot v_k)$ is built from the right and left eigenvectors $v_k$, $u_k$ and the operator-dressed boundary
vectors $\ell_O$, $r_O$. Each mode thus defines a correlation length and, for complex $\lambda_k=|\lambda_k|e^{i\phi_k}$, a wave vector,
\begin{equation}
\xi_k=-\frac{2}{\ln|\lambda_k|},\qquad q_k=\pi-\frac{|\phi_k|}{2},
\label{eq:xiq}
\end{equation}
where the branch of $q$ appropriate for $q\in[\pi/2,\pi]$ has been chosen. Because of U(1) symmetry, $\mathcal{T}$ is block diagonal in the
charge difference $\Delta S^z$ of its bra and ket indices. The sector $\Delta S^z=\pm1$ contains only spin-flip modes, while the sector
$\Delta S^z=0$ contains the identity, the $S^z$ components of triplet modes, singlet (dimer) modes, and modes that couple to neither. We
classify every mode of the $\Delta S^z=0$ sector by its weights for $O=S^z$ and for the bond energy $O=h=\Sv_0\cdot\Sv_1$: a mode is
\emph{triplet} if its normalized $S^z$ weight dominates, \emph{singlet} if its $h$ weight dominates, and \emph{dark} if both are below
$10^{-7}$ (Appendix~\ref{app:channels}). A label-free classification uses SU(2) symmetry: a spin-$S$ multiplet of the transfer matrix appears with the
same eigenvalue in the sectors $\Delta S^z=-S,\dots,S$. The leading $\Delta S^z=0$ eigenvalue without a partner in the $\Delta S^z=\pm1$ sector defines the
dimer (singlet) length $\xi_s$. The $\Delta S^z=\pm1$ sector contains all vector (rank-1) modes, i.e.\ besides the spin also the vector chirality
$\Sv_i\times\Sv_j$; its leading eigenvalue defines the longest triplet length $\xi_t$. Which operator the leading triplet mode belongs to is read off from
its weights for $S^z$ and for the vector chirality (Appendix~\ref{app:channels}). We use the symmetry-based lengths for all correlation lengths quoted
below; the weights serve to interpret amplitudes.

\paragraph*{Disorder point.}
Below the disorder point the leading triplet mode is real and positive, $q=\pi$; above it, the leading mode is a complex-conjugate pair and
$q<\pi$~\cite{Stephenson1970,Schollwoeck1996}. We monitor the ratio $|\lambda_{\mathrm{pair}}|/|\lambda_1|$ of the largest complex pair to the
largest triplet mode; the decay is incommensurate when this ratio is $1$.

\paragraph*{Finite-$\chi$ correlation lengths.}
At finite bond dimension $\chi$ the transfer-matrix spectrum is discrete, while the spectrum of the dimerized phase is a two-spinon continuum.
The leading eigenvalue therefore approaches the continuum edge from below, and $\xi(\chi)$ underestimates $\xi$. We extrapolate
$m_1=1/\xi$ linearly in the gap $\Delta m=m_2-m_1$ between the two leading distinct inverse lengths of the same channel,
$\Delta m\to0$~\cite{Rams2018}, using the three largest $\chi$ when $\Delta m$ decreases monotonically with $\chi$.

\subsection{Structure factor and the Lifshitz point}
\label{sec:Sk}

With the sublattice-averaged correlator $\bar C(r)=\tfrac12[C_0(r)+C_1(r)]$, the static structure factor of the translation-averaged state is
\begin{equation}
S(k)=\tfrac34+2\sum_{r\ge1}\bar C(r)\cos(kr),\qquad
S''(\pi)=-2\sum_{r\ge1}r^2(-1)^r\,\bar C(r).
\label{eq:Sk}
\end{equation}
For $\Jt<\JL$ the maximum is at $\pi$ and $S''(\pi)<0$; at the Lifshitz point the maximum splits, and $\JL$ is the root of $S''(\pi)$. This
definition uses only correlation functions, does not depend on a grid in $k$, and can be interpolated between values of $\Jt$. On a ring of $N$ sites,
$S(k)$ exists only at $k=2\pi m/N$, and we use either the discrete curvature $S(\pi+\tfrac{2\pi}N)+S(\pi-\tfrac{2\pi}N)-2S(\pi)$ or the ring version of
$S''(\pi)$, with $r$ taken in $(-N/2,N/2]$.

\section{Numerical methods}
\label{sec:methods}

\paragraph*{iDMRG.}
Ground states are obtained with the two-site iDMRG algorithm~\cite{White1992,McCulloch2008,Schollwoeck2011} as implemented in
TeNPy~\cite{TeNPy2018,TeNPy2024}, with a two-site unit cell, conservation of $S^z_{\rm tot}$, a subspace-expansion mixer during a
gradual increase of $\chi$, and a singular-value cutoff $10^{-12}$ ($10^{-14}$ near the MG point). For the dimerized phase
($0.38\le\Jt\le0.70$, 81 values, non-uniformly spaced with refinements near the MG and Lifshitz points) we use $\chi=48$, $64$, $96$, $128$, $192$ and $256$; correlation functions are computed up to
$r_{\max}=\max(60,12\xi)$. Near the MG point we use 25 values of $\Jt$ between $0.498$ and $0.510$, as close as $5\times10^{-5}$ to $\tfrac12$,
with $\chi=64$, $128$, $192$. For the Lifshitz point we use a grid of spacing $10^{-4}$
around the root with $\chi=64$ and $128$. On the gapless side we use 13 values $0\le\Jt\le0.30$ with $\chi=96$, $128$, $160$ and $192$. The channel
decomposition uses $\chi=128$ at 17 values $0\le\Jt\le0.70$ with blocks $l\le3$ and $d\le16$; at $0.32\le\Jt\le0.45$ it is extended to $d\le60$
($\chi=128$, $192$, $256$) and $d\le200$ ($\chi=128$), and at $\Jc$ and $\Jt=0$ to $l\le6$ ($\chi=192$). The split of the scalar channel into bond-energy and
chirality parts uses the stored states of the largest bond dimension. In all fits we state the number of points and free parameters; $\chi^2$ denotes
the usual fit statistic.

\paragraph*{Convergence at criticality.}
At $\Jt\le\Jc$ the iDMRG change of the half-chain entropy between sweeps, $\Delta S$, does not fall below $10^{-5}$--$10^{-4}$ even when the
observables of interest are converged. We therefore monitor the block mutual information itself (two probes, $I_3(3)$ and $I_2(4)$) every
100 sweeps and accept a state as \emph{converged} when the probes vary by less than $10^{-5}$ over three checkpoints, or as \emph{stable}
when they vary by less than $10^{-4}$ after at least 1500 sweeps; the remaining spread enters the error budget. A second, independent run
with a slower bond-dimension ramp and a stronger mixer is performed for states that drift and for gapless points that show a finite
dimerization; the lower-energy state is kept. For every $(\Jt,\chi)$ all available states are compared and states within $2\times10^{-8}$
of the lowest energy per site are treated as equivalent; the spread of their mutual information is an additional error (Appendix~\ref{app:convergence}).

\paragraph*{Finite-$\chi$ dimerization.}
Finite-bond-dimension (finite-$\chi$) iMPS with a two-site unit cell can develop a weak spontaneous dimerization at criticality, measured by the dimer
order parameter $D=|\langle h_0\rangle-\langle h_1\rangle|$. We assess its effect in three ways: by comparing dimerized and undimerized states of equal
bond dimension, by an explicit bond alternation $J_1\to J_1[1+\eta(-1)^i]$ with $\eta\le2\times10^{-3}$ at $\chi=128$, $160$, $192$
(Appendix~\ref{app:alternation}), and by periodic rings, whose ground states have $D=0$ by symmetry.

\paragraph*{Block reduced density matrices.}
For a right-canonical iMPS the left environment of site $s_0$ is $\mathrm{diag}(\Lambda^2)$ and the right environment of any site is the
identity. We build the open left block $T[a,a',\sigma,\sigma']$ of $A_1$, propagate it through the $d$ traced sites, and contract it with an
open right block of $A_2$ built from the identity; the latter depends only on the sublattice of $A_2$ and is computed once. The cost is
$O(d\,\chi^3 4^l+\chi^2 16^l)$, compared with $O(d\,\chi^3 16^l)$ for direct contraction, and blocks up to $l=6$ are accessible
(Appendix~\ref{app:contraction}). All block entropies agree with an independent TeNPy routine to $4\times10^{-15}$. For the channel decomposition the
SU(2) selection rule is satisfied to $4\times10^{-4}$ (relative weight of rank-mismatched components; $10^{-9}$ for most states).

\paragraph*{Large-$\chi$ values of the infinite chain.}
For $\chi\ge128$ the ratio $R$ of the iMPS is flat within $\sim10^{-3}$, whereas the largest correlation length $\xichi$ of the iMPS is not monotonic in
$\chi$ at criticality. These plateau values, however, lie systematically below the infinite-size limit obtained from periodic rings: at $\Jc$ by
$0.0005$--$0.0018$ ($x=1/4$) and $0.0013$--$0.0050$ ($x=1/9$), at $\Jt=0$ by $0.0024$--$0.0046$ and $0.0049$--$0.0096$, for $l=2$--$6$. A linear
extrapolation in $1/\xichi$ agrees with the rings within its larger errors (Appendix~\ref{app:convergence}). All quantitative statements on the
gapless side and at the KT point are therefore based on rings; iMPS are used for the dimerized phase and for the channel decomposition at
$d\le16$.

\paragraph*{Exact diagonalization and finite chains.}
Level spectroscopy uses exact diagonalization of periodic rings with $N=12$, $16$, $20$, $24$ at 31 values $0\le\Jt\le0.30$, in the $S^z=0$ and $S^z=1$
sectors; singlet and triplet levels are separated by comparing the two sectors. To relate the Lifshitz point to finite-size estimates we use (i)
open chains of $N=40$--$320$ sites with finite DMRG ($\chi=64$, $0.517\le\Jt\le0.527$), with the full-chain structure factor
$S(k)=N^{-1}\sum_{ij}e^{ik(i-j)}\langle\Sv_i\cdot\Sv_j\rangle$ or the bulk correlator of the two central sites, and (ii) periodic rings, by exact
diagonalization for $N\le24$ and by DMRG for $N=32$, $48$, $64$ ($\chi=160$), at $0.47\le\Jt\le0.56$. On a ring the ground state is a
translation-invariant superposition of the two dimer patterns whose interference contributes $O(2^{-N/2})$ to $C(r)$; we therefore also compute the
dimer partner state (the lowest excited singlet, obtained in DMRG as the lowest state orthogonal to the ground state) and use the average of the two
correlators, which equals the sublattice-averaged correlator of a symmetry-broken state. Near the Lifshitz point the DMRG rings $N=48$, $64$ are also
computed on a grid of spacing $10^{-4}$, with $\chi=160$, and for $N=48$ also with $\chi=240$ and $320$.

\paragraph*{Periodic rings for block mutual information.}
Block mutual information on rings is computed with finite DMRG on periodic chains (TeNPy, two-site, $S^z_{\rm tot}$ conserved), with the blocks centered
in the matrix-product state and two placements shifted by one site averaged; their difference enters the error. We use rings of $N=32$, $48$, $64$ at
$\Jt=0$, $0.10$--$0.22$ (step $0.02$), $0.23$, $\Jc$ and $0.25$--$0.28$ (step $0.01$), and in addition $N=80$ and $96$ at $\Jc$, with two or three bond
dimensions per ring: $\chi=192$, $256$ for $N=32$; $256$, $320$, $384$ for $N=48$, $64$; $256$, $320$ for $N=80$; $384$, $448$ for $N=96$. For $l=6$ the
bond dimension is limited to $\chi=320$ by memory. Geometries are $l=2$--$6$ with $d=l/2$, $l$, $2l$ ($x=4/9$, $1/4$, $1/9$) and, for $\Jt\ge0.20$, $d=3l$
($x=1/16$); at $\Jc$ we add for every ring the largest possible gap (blocks on opposite sides of the ring), which gives the smallest cross ratio a ring
allows. On a ring the cross ratio is built from chord lengths, $x=\sin^2(\pi l/N)/\sin^2[\pi(l+d)/N]$, for which the CFT prediction is unchanged.
Channel shares are computed for $l\le6$ at $\Jc$ and for $l\le4$ at the other couplings. All values are extrapolated to $N\to\infty$ at fixed $l$ and
$d$; the extrapolation and the error budget, which contains the bond dimension, the ring size, the placement and the DMRG convergence, are described in
Appendix~\ref{app:rings}. Periodic DMRG reproduces exact diagonalization of a ring of $16$ sites at $\Jt=0.26$ to $3.5\times10^{-14}$ in the mutual
information.

\section{Results and discussion}
\label{sec:results}
Throughout this section, we use $J_1=1$, so that $\Jt$ denotes the dimensionless ratio $\Jt/J_1$.

\subsection{Correlation lengths and the two correlation channels}
\label{sec:xi}

\begin{figure*}[t]
\centering
\figfile{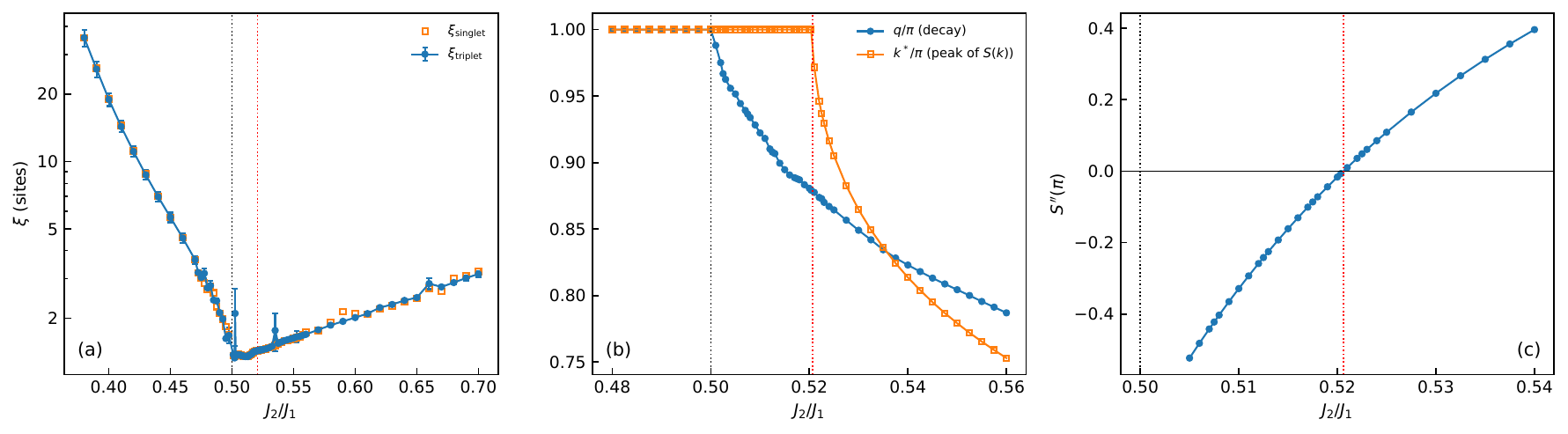}{\textwidth}
\caption{Commensurate--incommensurate crossover in the dimerized phase (infinite chain). Dotted lines: MG point (black) and $\JL=0.520622$
(red). (a) Triplet (filled) and singlet (open) correlation lengths from the weight-based mode labels; the triplet length is extrapolated in
the transfer-matrix gap. Table~\ref{tab:trackA} (symmetry-based labels) differs at $\Jt=0.495$ and by $2$--$3\%$ for $0.515\le\Jt\le0.53$.
(b) Wave vector $q/\pi$ of the leading complex transfer-matrix pair (circles) and peak position $k^*/\pi$ of $S(k)$ (squares); between the
dotted lines the decay is incommensurate while $S(k)$ still peaks at $\pi$. (c) $S''(\pi)$ [Eq.~\eqref{eq:Sk}]; its root defines $\JL$.}
\label{fig:onsets}
\end{figure*}

Table~\ref{tab:trackA} and Fig.~\ref{fig:onsets}(a) summarize the correlation lengths in the dimerized phase. The triplet length $\xi_t$ is taken from
the leading eigenvalue of the $\Delta S^z=\pm1$ sector of the transfer matrix, the dimer (singlet) length from the leading $\Delta S^z=0$ eigenvalue
without a partner in that sector (Sec.~\ref{sec:TM}). The $\Delta S^z=\pm1$ sector contains the vector modes, spin and vector chirality. For
$\Jt\lesssim0.50$ its leading mode is the spin mode. For $\Jt\ge0.51$ the leading mode is a vector-chirality mode in 47 of 48 states at $\chi=256$; its
length differs from that of the leading spin mode by $-1$ to $+5\%$, e.g.\ $1.406$ against $1.376$ at $\Jt=0.515$ and $1.467$ against $1.430$ at $0.52$.
This explains why $\xi_t$ differs by $2$--$3\%$ from the weight-based spin length for $0.515\le\Jt\le0.53$ (Appendix~\ref{app:channels}); at $\Jt=0.495$
the symmetry-based identification removes an ambiguity between nearly degenerate modes ($1.838$ instead of $1.625$). The extrapolation in the transfer-matrix gap matters for $\Jt\lesssim0.45$: at $\Jt=0.40$ it adds 13\% to the largest-$\chi$ value
($\xi=18.9\pm1.2$), and at $\Jt=0.38$ we find $\xi=35.6\pm3.1$. Close to and above the MG point $\xi$ is short and converged at the largest $\chi$. It has
a shallow minimum $\xi=1.35\pm0.02$ between $\Jt=0.502$ and $0.511$, where the scatter reflects switches between discrete transfer-matrix modes, and
grows again in the incommensurate regime. The minimum lies about $6\%$ below the value $\xi\simeq1.44$ obtained as $\Jt\to\tfrac12$
(Sec.~\ref{sec:MG}); at $\Jt=\tfrac12$ itself the correlations beyond one bond vanish because their amplitude, not their decay rate, goes to zero.

The singlet (dimer) and triplet correlation lengths agree: for the 80 states of the dimerized-phase grid at $\chi=256$ the median ratio is
$\xi_s/\xi_t=0.9996$, with $10$--$90\%$ of the values within $0.989$--$1.007$; relative to the spin mode alone the median is $0.994$ ($10$--$90\%$:
$0.988$--$1.001$). The largest deviations ($0.974$ at $\Jt=0.38$) occur where $\xi$ is long and not yet converged in $\chi$. The measured correlators confirm this without the transfer matrix: the envelope decay rates of the dimer--dimer correlator and
of $C(r)$ over the largest reliable distances have a median ratio of $1.02$ ($10$--$90\%$: $0.82$--$1.21$). This is the expected behavior if both
correlation functions are dominated by two-spinon states: the lowest singlet and triplet two-spinon states share the same continuum edge, so the
asymptotic decay of spin and dimer correlations is set by the same branch point. In the $J_1$--$J_2$ chain, block mutual information therefore reveals no
additional asymptotic correlation length beyond the common spin and dimer scale; whatever it adds beyond the spin correlator must appear in amplitudes,
which we analyze in Sec.~\ref{sec:channelresults}.

\begin{table*}[t]
\caption{Correlations in the dimerized phase (infinite chain). $\xi_t$: longest triplet (rank-1) correlation length ($\Delta S^z=\pm1$ sector),
extrapolated in the transfer-matrix gap~\cite{Rams2018}; for $\Jt\ge0.51$ it belongs to a vector-chirality mode and differs from the spin length by
$-1$ to $+5\%$ (Sec.~\ref{sec:xi}). $^\dagger$: unextrapolated $\chi=256$ values (uncertainty $\approx0.02$ from mode switching for
$0.501\le\Jt\le0.515$). $\xi_s/\xi_t$: dimer (singlet) to triplet length ratio at $\chi=256$, both identified by symmetry. $q$: wave vector of the
leading complex pair ($\pi$ if the leading mode is real). $k^*$: maximum of $S(k)$; $S''(\pi)$: Eq.~\eqref{eq:Sk}; $D=|\langle h_0\rangle-\langle h_1\rangle|$:
dimer order parameter. At $\Jt=\tfrac12$ all correlations beyond one bond vanish.}
\label{tab:trackA}
\begin{ruledtabular}
\begin{tabular}{lcccccc}
$\Jt$ & $\xi_t$ & $\xi_s/\xi_t$ & $q/\pi$ & $k^*/\pi$ & $S''(\pi)$ & $D$ \\
\hline
0.38    & $35.6\pm3.1$      & 0.974 & 1       & 1      & $-233.8$  & 0.421 \\
0.40    & $18.9\pm1.2$      & 0.985 & 1       & 1      & $-70.34$  & 0.507 \\
0.42    & $11.08\pm0.58$    & 0.989 & 1       & 1      & $-26.12$  & 0.579 \\
0.45    & $5.64\pm0.28$     & 0.991 & 1       & 1      & $-7.321$  & 0.663 \\
0.47    & $3.65\pm0.14$     & 1.000 & 1       & 1      & $-3.262$  & 0.705 \\
0.48    & $2.733\pm0.015$   & 1.003 & 1       & 1      & $-2.128$  & 0.722 \\
0.49    & $2.119\pm0.002$   & 1.000 & 1       & 1      & $-1.326$  & 0.737 \\
0.495   & $1.838^\dagger$   & 1.002 & 1       & 1      & $-1.015$  & 0.744 \\
0.50    & --                & --    & --      & 1      & $-0.750$  & 0.750 \\
0.501   & $1.376^\dagger$   & 1.000 & 0.988   & 1      & $-0.702$  & 0.751 \\
0.505   & $1.386^\dagger$   & 0.997 & 0.952   & 1      & $-0.523$  & 0.756 \\
0.51    & $1.360^\dagger$   & 1.002 & 0.922   & 1      & $-0.329$  & 0.761 \\
0.515   & $1.406^\dagger$   & 1.022 & 0.895   & 1      & $-0.161$  & 0.765 \\
0.52    & $1.467^\dagger$   & 0.999 & 0.881   & 1      & $-0.0166$ & 0.769 \\
0.525   & $1.482^\dagger$   & 0.999 & 0.864   & 0.905  & $0.109$   & 0.773 \\
0.53    & $1.500^\dagger$   & 1.000 & 0.849   & 0.865  & $0.218$   & 0.777 \\
0.54    & $1.578^\dagger$   & 0.994 & 0.823   & 0.814  & $0.396$   & 0.782 \\
0.55    & $1.647^\dagger$   & 0.992 & 0.804   & 0.779  & $0.533$   & 0.786 \\
0.60    & $2.015\pm0.004$   & 0.996 & 0.733   & 0.688  & $0.891$   & 0.788 \\
0.65    & $2.47\pm0.04$     & 1.000 & 0.687   & 0.645  & $1.040$   & 0.769 \\
0.70    & $3.15\pm0.09$     & 0.989 & 0.654   & 0.619  & $1.147$   & 0.735 \\
\end{tabular}
\end{ruledtabular}
\end{table*}

\subsection{Vicinity of the Majumdar--Ghosh point and the disorder point}
\label{sec:MG}

\begin{figure}[t]
\centering
\figfile{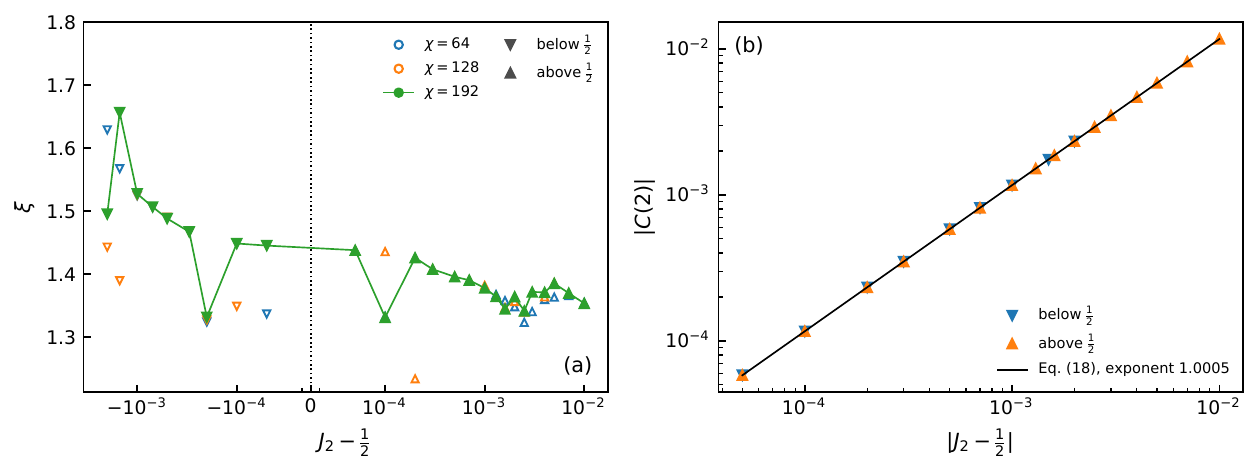}{\columnwidth}
\caption{Vicinity of the MG point; triangles down (up): $\Jt<\tfrac12$ ($>\tfrac12$). (a) Leading triplet correlation length versus
$\Jt-\tfrac12$ (symmetric log axis) for $\chi=192$ (filled) and $64$, $128$ (open); $\xi$ is continuous at $\tfrac12$ ($\simeq1.44$). Outliers
and the spread at $\Jt-\tfrac12\le-1.5\times10^{-3}$ are mode switches. (b) $|C(2)|$ versus $|\Jt-\tfrac12|$ ($\chi=192$); line:
Eq.~\eqref{eq:C2}.}
\label{fig:MG}
\end{figure}

Exactly at $\Jt=\tfrac12$ the iMPS has bond dimension 2 and all correlations beyond the dimer vanish. Figure~\ref{fig:MG} shows the approach to
this point from both sides. The spin correlator at distance two vanishes linearly,
\begin{equation}
|C(2)|=1.169\,|\Jt-\tfrac12|^{1.0005},
\label{eq:C2}
\end{equation}
over two decades in $|\Jt-\tfrac12|$ ($5\times10^{-5}$ to $10^{-2}$), on both sides, and the correlation length is continuous: $\xi=1.4455$ at
$\Jt-\tfrac12=-5\times10^{-5}$ and $\xi=1.4382$ at $+5\times10^{-5}$, identical for $\chi=128$ and $192$ (Table~\ref{tab:MG}).

\begin{figure}[t]
\centering
\figfile{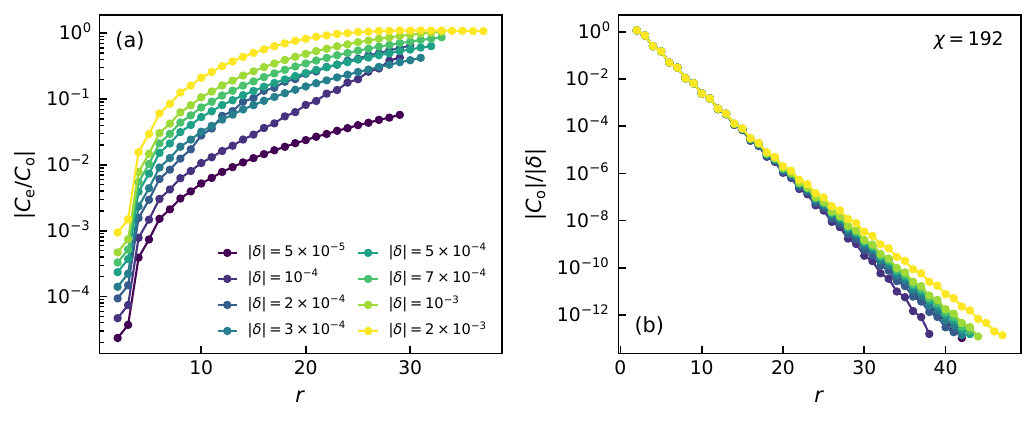}{\columnwidth}
\caption{Real-space structure around the MG point ($\chi=192$), $\delta=\Jt-\tfrac12$, with the even and odd parts
$C_{\rm e,o}(r)=\tfrac12[C(r;\delta)\pm C(r;-\delta)]$ of the spin correlator. (a) $|C_{\rm e}/C_{\rm o}|$ versus distance.
(b) $|C_{\rm o}|/|\delta|$, which to first order in $\delta$ is the correlation profile $|\Phi_1(r)|$; the curves for all $\delta$ agree within
$5\%$ up to $r=10$ and separate gradually at larger $r$.}
\label{fig:MGrealspace}
\end{figure}

\paragraph*{Real-space structure.}
Writing $\delta=\Jt-\tfrac12$, perturbation theory about the
dimer product state gives
$C(r;\delta)=\delta\,\Phi_1(r)+\delta^2\Phi_2(r)+O(\delta^3)$,
where the first-order profile $\Phi_1(r)$ is fixed by the
excitations of the MG Hamiltonian and is the same on both sides.
The even and odd parts
$C_{\rm e,o}(r)=\tfrac12[C(r;\delta)\pm C(r;-\delta)]$
isolate the two orders directly in real space, without the
transfer matrix [Fig.~\ref{fig:MGrealspace}].
For eight values $5\times10^{-5}\le|\delta|\le2\times10^{-3}$,
the odd part divided by $|\delta|$ collapses onto a single
curve, within $5\%$ over the first three decades of decay ($r\le10$);
its decay length approaches $\xi\simeq1.44$ at large $r$.

The ratio $|C_{\rm e}/C_{\rm o}|$ equals $0.469\,|\delta|$
at $r=2$ for all $\delta$, as required for a regular expansion,
and grows with $r$. At fixed $r$ it is roughly proportional
to $|\delta|$ (at $r=10$, from $0.005$ for
$|\delta|=5\times10^{-5}$ to $0.21$ for
$|\delta|=2\times10^{-3}$), as expected for a second-order
correction. The even part thus becomes progressively more
important in the long-distance tail, but its size there
depends strongly on $|\delta|$: at $r=29$ it is $6\%$
of the odd part for $|\delta|=5\times10^{-5}$ and close
to unity for $|\delta|\ge10^{-3}$. 
The correlators on opposite sides of the MG point therefore
develop different long-distance profiles, without a common,
sharply defined crossover distance. Since their first-order
profiles differ only by an overall sign, the asymmetry beyond
this sign reversal arises at second order in $\delta$ and
becomes most apparent in the long-distance tail. This
behavior is consistent with the onset of incommensurate
decay identified independently by the transfer-matrix
analysis below.

\paragraph*{Transfer-matrix onset.}
The complex pair takes over the leading modulus, i.e.\ the decay becomes incommensurate, between $\Jt=0.5003$ and $0.5005$ for all three bond
dimensions, and stays leading for all larger $\Jt$. The disorder point thus coincides with the MG point to within $5\times10^{-4}$, in agreement with
earlier work~\cite{WhiteAffleck1996,Bursill1995}. Near $\tfrac12$ the state is close to a product state, and the bond dimension actually used (24--64 for
$\chi=64$, 24--75 for $\chi\ge128$) is set by its Schmidt spectrum rather than by the nominal $\chi$. Varying the singular-value cutoff between $10^{-10}$
and $10^{-14}$ changes neither the bond dimension used nor $\xi_t$ by more than $10^{-3}$, so the values near $\tfrac12$ are converged. Above the onset
$\pi-q$ grows continuously, but its onset law cannot be determined with the present method: the finite transfer-matrix spectrum discretizes a two-spinon
continuum with few modes, and the leading mode switches between discrete members of this continuum, producing a jump of $\pi-q$ near
$\Jt-\tfrac12\simeq0.0018$ (Appendix~\ref{app:channels}). More fundamentally, close to $\tfrac12$ the modulation wavelength $\pi/(\pi-q)$ exceeds $10^2$
sites while $\xi\simeq1.4$, so the correlation functions carry no measurable information about $q$; only their analytic structure does.

\begin{table}[t]
\caption{Vicinity of the MG point. Onset bracket of the incommensurate decay and $\xi$ at the two grid points closest to $\tfrac12$
($\Jt-\tfrac12=\mp5\times10^{-5}$). The range of bond dimensions actually used is set by the Schmidt spectrum of the nearly product state (cutoff $10^{-14}$; unchanged for cutoffs $10^{-10}$--$10^{-14}$). The
$\chi=64$ value below $\tfrac12$ is affected by a mode switch.}
\label{tab:MG}
\begin{ruledtabular}
\begin{tabular}{lcccc}
$\chi$ & onset bracket & $\chi_{\rm used}$ & $\xi(\tfrac12^-)$ & $\xi(\tfrac12^+)$ \\
\hline
64  & (0.5003, 0.5005] & 24--64 & (1.3368) & 1.4382 \\
128 & (0.5003, 0.5005] & 24--75 & 1.4454 & 1.4382 \\
192 & (0.5003, 0.5005] & 24--75 & 1.4455 & 1.4382 \\
\end{tabular}
\end{ruledtabular}
\end{table}

\subsection{The Lifshitz point}
\label{sec:Lifshitz}

The curvature $S''(\pi)$ is a smooth function of $\Jt$ [Fig.~\ref{fig:onsets}(c)] with a simple root. On a grid of spacing $10^{-4}$ we find
$\JL=0.5206218$ ($\chi=64$) and $0.5206226$ ($\chi=128$); linear and quadratic interpolation of the root agree to $10^{-6}$. We combine this
bond-dimension and interpolation uncertainty with the largest deviation of the independent finite-size routes discussed below,
\begin{equation}
\JL=0.520622\pm0.000001\,(\chi,\ \mathrm{interpolation})\pm0.000002\,(\mathrm{finite\ size})=0.520622(2).
\label{eq:JL}
\end{equation}
The root does not depend on whether $S(k)$ is built from the sublattice-averaged correlator or from correlations centered on one site.
Consistently, $k^*$ leaves $\pi$ between $\Jt=0.52$ and $0.525$ (Table~\ref{tab:trackA}).

\begin{figure*}[t]
\centering
\figfile{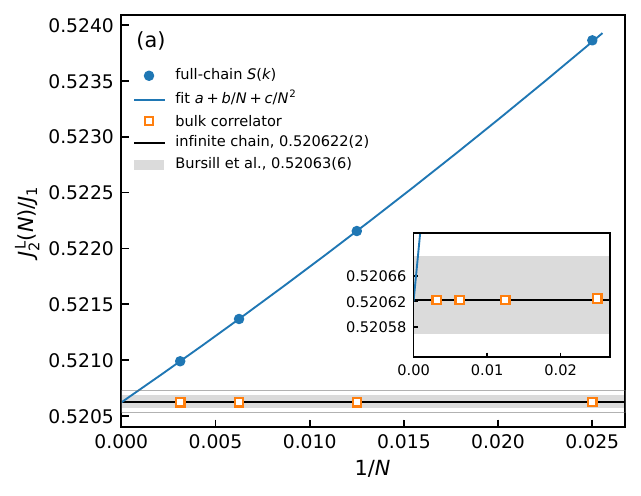}{0.48\textwidth}\hfill
\figfile{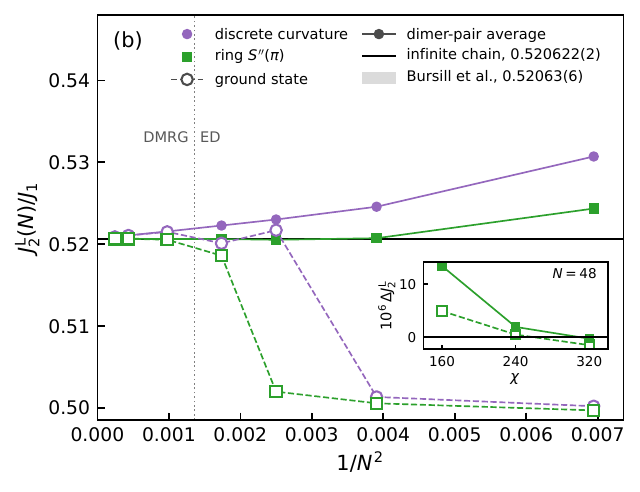}{0.48\textwidth}
\caption{Finite-size estimates of the Lifshitz point. (a) Open chains versus $1/N$: full-chain structure factor (circles) with
Eq.~\eqref{eq:JLN} and bulk correlator (open squares); inset: magnification of the region of the bulk values.
(b) Periodic rings versus $1/N^2$: ED ($N\le24$) and DMRG ($N=32$--$64$, $\chi=160$, grid spacing $0.0025$); circles, discrete
curvature; squares, ring $S''(\pi)$; open symbols and dashed lines, ground state only; filled symbols and solid lines, dimer-pair
average. Inset: ring $S''(\pi)$ root at $N=48$ (grid spacing $10^{-4}$) minus the infinite-chain root, versus $\chi$. Black line:
Eq.~(19); gray band: value $0.52063(6)$ of Ref.~\cite{Bursill1995}.}
\label{fig:Lifshitz}
\end{figure*}

Equation~\eqref{eq:JL} agrees with the value $0.52063(6)$ obtained by Bursill \emph{et al.}~\cite{Bursill1995} with infinite-system DMRG and
essentially the same criterion, the vanishing second derivative of the Fourier-transformed correlator at $\pi$; the same value is tabulated in
Ref.~\cite{RothSchollwoeck1998}. Some later works quote $0.52036(6)$ with reference to Ref.~\cite{Bursill1995}; this appears to be a transposition of
the last two digits. Our calculation thus independently confirms the original estimate with a smaller uncertainty, using translation-invariant
states. Table~\ref{tab:Lifshitz} and Fig.~\ref{fig:Lifshitz} show how finite-size estimates approach it.

\paragraph*{Open chains.}
The bulk correlator of the two central sites reproduces Eq.~\eqref{eq:JL} to $10^{-6}$ for $N\ge80$. The full-chain structure factor contains boundary
contributions, and its root moves as
\begin{equation}
\JL(N)=0.520621+\frac{0.1163}{N}+\frac{0.515}{N^2},
\label{eq:JLN}
\end{equation}
obtained by solving for the three coefficients with $N=80$, $160$, $320$ (no degree of freedom remains); it predicts $0.523850$ for $N=40$, compared with
the computed $0.523863$. Two-size extrapolations linear in $1/N$ converge slowly: $0.52045$ from $N=40$ and $80$, $0.52058$ from $80$ and $160$.

\paragraph*{Periodic rings.}
On a ring the ground state is a translation-invariant superposition of the two dimer patterns. For $N=48$ and $64$ (DMRG, $\chi=160$, grid spacing
$10^{-4}$) the root of the ring version of $S''(\pi)$ is $0.520628$ and $0.520629$ from the ground state alone and $0.520636$ for both sizes from the
dimer-pair average. The offset of $0.5$--$1.3\times10^{-5}$ from Eq.~\eqref{eq:JL}, the same for both ring sizes, is a truncation effect, not a
finite-size effect: for $N=48$ the root moves to within $1.9\times10^{-6}$ of Eq.~\eqref{eq:JL} at $\chi=240$ and within $1.5\times10^{-6}$ at $\chi=320$,
for both definitions, with a change of sign between the two bond dimensions. Correspondingly, the ring correlator at $6\le r\le16$ differs from the
infinite-chain correlator by up to $2.6\times10^{-7}$ at $\chi=160$ but by less than $3.6\times10^{-8}$ at $\chi=240$ and $320$. (On a coarser grid,
$\Delta\Jt=0.0025$, linear interpolation had shifted the roots by a further $3$--$4\times10^{-5}$.) The discrete curvature of $S(k)$ at $\pi$ converges
more slowly ($0.52154$, $0.52106$, $0.52088$ for $N=32$, $48$, $64$; $0.52066$ extrapolated). Small rings behave very differently: for $N\le16$ the
ground state alone gives $0.500$--$0.501$, because the interference between the two dimer patterns dominates the small correlations near $\tfrac12$, and
extrapolations of the dimer-pair average from $N\le24$ scatter between $0.5175$ and $0.5211$ depending on the definition and the assumed form.

\begin{table}[t]
\caption{Lifshitz point. Infinite chain (iDMRG, root of $S''(\pi)$); open chains (finite DMRG, $\chi=64$) with the full-chain structure factor or the
bulk correlator; periodic rings (DMRG, grid spacing $10^{-4}$) with the ring version of $S''(\pi)$, from the ground state (gs) or the dimer-pair
average (pair).}
\label{tab:Lifshitz}
\begin{ruledtabular}
\begin{tabular}{llc}
System & definition & $\JL$ \\
\hline
infinite, $\chi=64$, $128$  & $S''(\pi)$, Eq.~\eqref{eq:Sk} & 0.5206218, 0.5206226 \\
open, $N=80$--$320$ & bulk & 0.520622 \\
open, $N=40,80,160,320$ & full chain & 0.523863, 0.522156, 0.521368, 0.520990 \\
open, $N\to\infty$ & full chain, Eq.~\eqref{eq:JLN} & 0.520621 \\
open, $N=40,80$ / $80,160$ & full chain, linear in $1/N$ & 0.52045 / 0.52058 \\
ring, $N=48$, $64$, $\chi=160$ & ring $S''(\pi)$, gs / pair & 0.520628, 0.520629 / 0.520636, 0.520636 \\
ring, $N=48$, $\chi=240$ & ring $S''(\pi)$, gs / pair & 0.520623 / 0.520624 \\
ring, $N=48$, $\chi=320$ & ring $S''(\pi)$, gs / pair & 0.520621 / 0.520622 \\
earlier work, infinite-system DMRG~\cite{Bursill1995} & second derivative at $\pi$ & 0.52063(6) \\
\end{tabular}
\end{ruledtabular}
\end{table}

\paragraph*{Three regimes.}
Taken together, the dimerized phase between $\Jc$ and $0.7$ contains three regimes for the short-range order:
(i) $\Jc<\Jt<\tfrac12$, commensurate decay and $S(k)$ peaked at $\pi$; (ii) $\tfrac12<\Jt<0.520622$, incommensurate decay but $S(k)$ still
peaked at $\pi$; (iii) $\Jt>0.520622$, incommensurate decay and a split peak. In regime (ii) the answer to ``is the state commensurate?''
depends on the probe: real-space quantities, including the mutual information, already see the incommensurate wave vector, whereas
the structure factor does not.

\subsection{Mutual information in the dimerized phase}
\label{sec:MIdimer}

For all states the closed form~\eqref{eq:fC} agrees with the directly computed two-site mutual information to machine precision ($10^{-15}$ in
all tests). In the incommensurate regime $I(r)=f(C(r))$ has nodes wherever $C(r)$ changes sign, with spacing $\pi/q$, i.e.\ wave vector $2q$.

\begin{figure*}[t]
\centering
\figfile{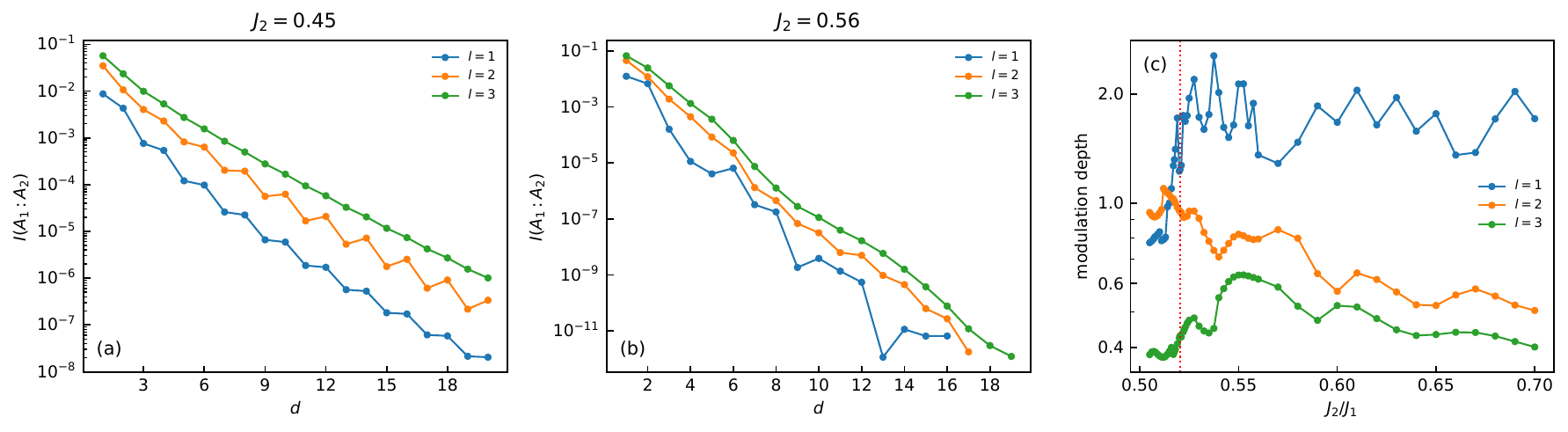}{\textwidth}
\caption{Block mutual information in the dimerized phase ($\chi=192$, averaged over the two sublattices).
(a) $I_l(d)$ at $\Jt=0.45$ (commensurate) and (b) at $\Jt=0.56$ (incommensurate) for $l=1,2,3$; values below $10^{-12}$ are numerical noise and omitted.
(c) Modulation depth of $\ln I_l(d)$ after removal of the exponential envelope, versus $\Jt$; the red dotted line marks $\JL$.}
\label{fig:blockMI}
\end{figure*}

Figure~\ref{fig:blockMI}(a,b) shows $I_l(d)$ for $l=1,2,3$ at a commensurate ($\Jt=0.45$) and an incommensurate ($\Jt=0.56$) point. All block sizes decay
with the same exponential rate, $2/\xi$. The local decay rate approaches this value only slowly, which indicates power-law prefactors. Fits of such
prefactors depend on the fit window even for the two-site quantity, for which $I=f(C)$ holds exactly, so we do not quote exponents.

To quantify the incommensurate modulation, we remove the exponential envelope, $\ln I_l(d)+2d/\xi$, subtract a linear trend, and use the root-mean-square residual as a measure of the modulation
depth. Relative to $l=1$ we find, over the 36 values $\Jt\ge0.52$ (median and interquartile range),
$\mathrm{mod}(l=2)/\mathrm{mod}(l=1)=0.44\,[0.34,0.53]$ and $\mathrm{mod}(l=3)/\mathrm{mod}(l=1)=0.28\,[0.25,0.34]$ [Fig.~\ref{fig:blockMI}(c)].
A block of $l$ sites sums correlations over $l^2$ site pairs whose phases $q(j-i)$ differ, so the nodes of individual $\cos^2$ terms are partially
filled in. Block mutual information is therefore a much smoother probe of the incommensurate regime than two-site quantities, whose nodes make
decay-length fits ill-conditioned.

\subsection{What block mutual information adds: SU(2) channels}
\label{sec:channelresults}

\begin{figure}[t]
\centering
\figfile{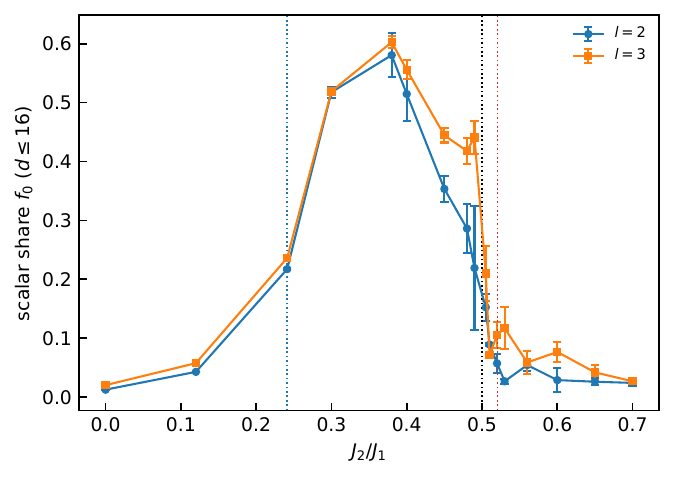}{0.6\columnwidth}
\caption{Share of block mutual information carried by the scalar (bond-energy, dimer) channel versus $\Jt$ for $l=2$ and $3$ at $d\le16$ ($\chi=128$).
Error bars: systematic uncertainty from the acceptance threshold of the second-order expansion ($5$--$15\%$) and the distance window. Dotted lines mark
$\Jc$ (blue), the MG point (black) and $\JL$ (red).}
\label{fig:channels}
\end{figure}

Table~\ref{tab:channels} and Fig.~\ref{fig:channels} give the share of block mutual information carried by the scalar channel for $d\le16$
[Eq.~\eqref{eq:channelMI}], with a systematic uncertainty from varying the acceptance threshold of the second-order expansion between $5\%$ and $15\%$
and the range of distances. The rank-2 channel carries at most $7\times10^{-3}$ and the rank-3 channel ($l=3$) at most $10^{-5}$ of the mutual
information at any valid point, so block mutual information consists of spin and scalar contributions only; components with different ranks on the two
blocks, forbidden by SU(2) symmetry, have a relative weight below $4\times10^{-4}$.

\paragraph*{Composition of the scalar channel.}
For $l=2$ the scalar channel contains only bond-energy correlations by construction. For $l=3$, the chirality part carries at most $2.4\times10^{-4}$ of
the mutual information for $\Jt\le0.45$, i.e.\ less than $0.1\%$ of the scalar channel, and the interference between bond energies and chirality is below
$0.5\%$ (below $2\%$ at $\Jt=0.48$--$0.49$) [Fig.~\ref{fig:scalarsplit}]. For the blocks studied here, the SU(2)-scalar channel is therefore dominated by
bond-energy correlations wherever its contribution is appreciable; we refer to it below as the dimer channel. Beyond the disorder point the chirality part
grows to $4$--$20\%$ of the scalar channel, which is then small, with a maximum of $2.7\%$ of the total mutual information just above $\tfrac12$. A natural
reading is that the incommensurate, spiral-like short-range correlations produce local chirality correlations; we have not analyzed this further.

\begin{figure*}[t]
\centering
\figfile{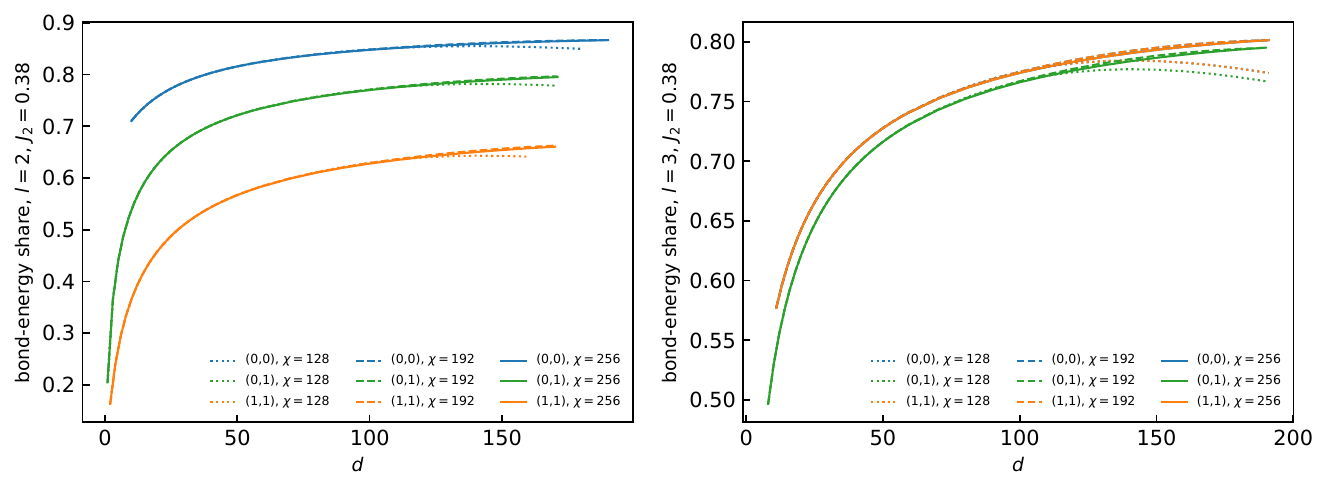}{\textwidth}
\caption{Bond-energy (dimer) share versus $d$ at $\Jt=0.38$ for $l=2$ (left) and $3$ (right); $\chi=128$ (dotted), $192$ (dashed), $256$ (solid).
$(s_1,s_2)$: sublattices of the first sites of the two blocks [$(0,0)$: both $l=2$ blocks on strong bonds; $(1,1)$: both on weak bonds]. The
large-$d$ decrease at $\chi=128$ is absent for $\chi\ge192$.}
\label{fig:chdist}
\end{figure*}

\paragraph*{Dependence on distance.}
In the commensurate dimerized regime the dimer share depends on the distance between the blocks over several correlation lengths, and for $l=2$ also on
the position of the blocks relative to the dimer pattern (Table~\ref{tab:chdist}, Fig.~\ref{fig:chdist}). At $\Jt=0.38$ ($\xi\approx36$) the share at the
largest distances computed, $d\approx170$--$190$, is $0.867$ when both blocks of $l=2$ cover strong bonds, $0.795$ for one strong and one weak bond, and
$0.661$ for two weak bonds; for $l=3$ the position matters little ($0.795$--$0.801$). At $\Jt=0.40$ ($\xi\approx19$) the corresponding values at
$d\approx90$--$110$ are $0.801$, $0.687$ and $0.466$ ($l=2$) and $0.70$--$0.71$ ($l=3$); at $\Jt=0.45$ ($\xi\approx5.6$) and $d\approx32$--$40$ they are $0.630$,
$0.394$, $0.124$ and $0.47$--$0.49$. The values for $\chi=192$ and $256$ agree within $1.5\times10^{-3}$, and the share is still increasing at the largest
distances computed, so these numbers are lower bounds for the long-distance shares. At $\chi=128$ the share passes through a maximum at $d\approx140$ and
then decreases; this decrease is absent at $\chi=192$ and $256$ and is therefore a finite-$\chi$ effect, consistent with the stronger finite-$\chi$
splitting of singlet and triplet lengths at small $\chi$. Since spin and dimer correlations share the same exponential decay, we attribute the rise to
different power-law prefactors of the two channels. Closer to the transition, at $\Jt=0.32$--$0.36$, the share exceeds $0.78$ at $d=60$ and is still rising.
Short-distance values ($d\le16$) underestimate the long-distance share in this regime.

\paragraph*{Dependence on $\Jt$.}
At fixed short distance ($d\le16$) the dimer share is small in the gapless phase ($1$--$2\%$ at $\Jt=0$, $4$--$6\%$ at $0.12$), reaches $22$--$24\%$ at
$\Jc$, exceeds one half for $0.30\lesssim\Jt\lesssim0.40$, decreases towards the MG point and falls to a few percent beyond the disorder point. In the
incommensurate regime the shares fluctuate with $d$ because of the nodes of the modulation, and only their average is meaningful. Block mutual information
thus contains no additional asymptotic correlation length (Sec.~\ref{sec:xi}), but in the commensurate dimerized phase up to $\Jt=0.40$ bond-energy correlations, which the
spin correlator does not contain, account for most of it at long distances. The dimer share decreases already within the commensurate regime and
reaches a few percent between the disorder and Lifshitz points, so the onset of incommensurate decay is accompanied by a change in an amplitude, not
only in a wave vector.

\begin{figure*}[t]
\centering
\figfile{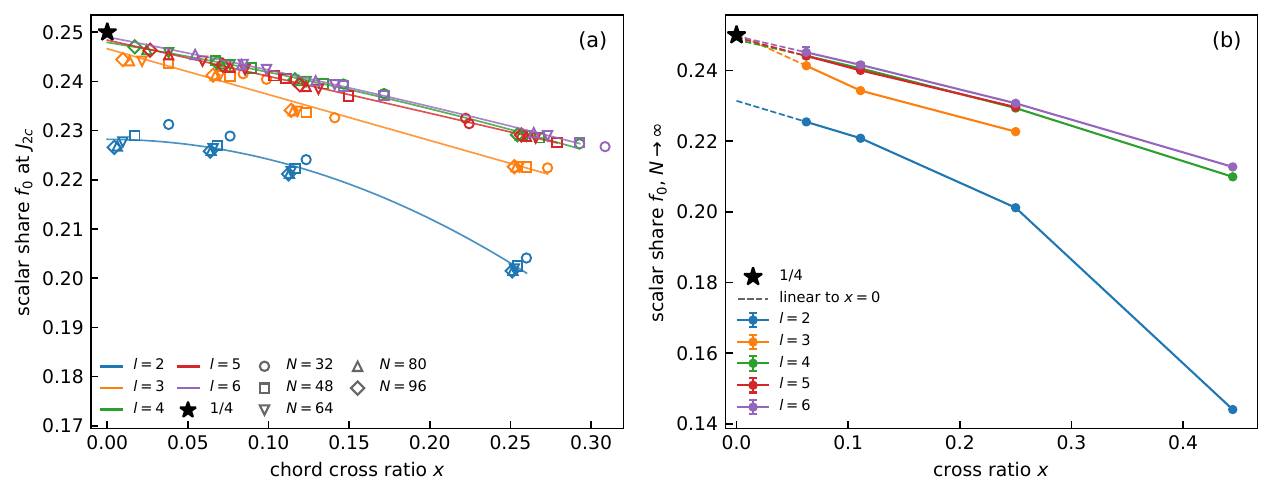}{\textwidth}
\caption{Scalar share at the KT point on periodic rings. (a) All rings and distances versus the chord cross ratio ($x\le0.32$, largest $\chi$
per ring; color: $l$, symbol: $N$), with quadratic fits for $x\le0.3$ extended to $x=0$. (b) $N\to\infty$ values at fixed geometry versus $x$;
dashed: linear extrapolation through $x=1/16$ and $1/9$.}
\label{fig:share}
\end{figure*}

\paragraph*{The critical point.}
At the SU(2)$_1$ fixed point the three components of the staggered magnetization and the dimerization are the four components of a single primary field
of scaling dimension $\tfrac12$, which are rotated into each other by the SO(4) symmetry of the fixed point~\cite{Affleck1989}. Their equal contributions
to the leading small-$x$ term of the mutual information [Eq.~\eqref{eq:Cardy}] imply a scalar share $1/4$ for large blocks at small $x$. Table~\ref{tab:quarter}
lists the scalar share at $\Jc$ on periodic rings, extrapolated to $N\to\infty$, for $x=4/9$, $1/4$, $1/9$ and $1/16$, and Fig.~\ref{fig:share} shows
it. At fixed $l$ the share increases as $x$ decreases. A linear extrapolation in $x$ through $x=1/16$ and $1/9$ gives $0.2504(1)$, $0.2487(3)$,
$0.2494(5)$ and $0.2497(36)$ for $l=3$--$6$, and the quadratic through $1/16$, $1/9$ and $1/4$ agrees within $0.0025$ ($l=3$) and $0.0004$ ($l\ge4$). A
second, independent route avoids the extrapolation in $N$. In a CFT the share on a ring depends on $N$ and $d$ only through the chord cross ratio, and
blocks on opposite sides of the largest rings reach $x=0.017$ ($l=4$), $0.027$ ($l=5$) and $0.054$ ($l=6$), where the share is $0.2470$, $0.2464$ and
$0.2455$. For $l\ge4$ the values of all rings and distances fall on one curve as a function of the chord cross ratio (root-mean-square deviation
$0.9$--$4.6\times10^{-4}$ from a quadratic in $x$ for $x\le0.3$), which extrapolates to $0.2479$, $0.2483$ and $0.2490$ for $l=4$, $5$ and $6$. For $l=2$
and $3$ the collapse is poorer (deviation $1.4$--$1.7\times10^{-3}$) and the share at $l=2$ stays near $0.23$. Both routes thus give a share within
$0.0021$ of $1/4$ at small $x$ for $l\ge4$, with the deviation decreasing with $l$. At $\Jt=0$, in contrast, the scalar share of the infinite chain at
$d=6l$ is only $0.014$--$0.031$ for $l=2$--$6$ and grows slowly with $l$, consistent with the different multiplicative logarithmic corrections that the
marginal coupling produces in the spin and dimer channels~\cite{Affleck1989}, which suppress the dimer channel as long as $g>0$. How the share changes at
first order in the coupling is discussed in Sec.~\ref{sec:CFTresults}.

\begin{table}[t]
\caption{Scalar (bond-energy, dimer) share $f_0$ of block mutual information for $l=2$ and $3$ ($\chi=128$) at the largest distances $d\le16$ where
the second-order decomposition is valid; uncertainties from the acceptance threshold ($5$--$15\%$) and the distance window. The spin channel carries
$1-f_0$. Last column: chirality-to-scalar ratio for $l=3$. For $\Jt\lesssim0.45$, $f_0$ still grows with $d$ (Table~\ref{tab:chdist}).}
\label{tab:channels}
\begin{ruledtabular}
\begin{tabular}{lccc}
$\Jt$ & $f_0(l=2)$ & $f_0(l=3)$ & chirality / scalar ($l=3$) \\
\hline
0.00  & $0.012\pm0.000$ & $0.020\pm0.001$ & $<0.1\%$ \\
0.12  & $0.043\pm0.001$ & $0.058\pm0.001$ & $<0.1\%$ \\
$\Jc$ & $0.217\pm0.002$ & $0.236\pm0.001$ & $<0.1\%$ \\
0.30  & $0.518\pm0.009$ & $0.519\pm0.004$ & $<0.1\%$ \\
0.38  & $0.581\pm0.038$ & $0.603\pm0.010$ & $<0.1\%$ \\
0.40  & $0.515\pm0.046$ & $0.556\pm0.016$ & $<0.1\%$ \\
0.45  & $0.353\pm0.022$ & $0.445\pm0.012$ & $<0.1\%$ \\
0.48  & $0.286\pm0.042$ & $0.418\pm0.022$ & $0.3\%$ \\
0.49  & $0.219\pm0.105$ & $0.441\pm0.028$ & $0.7\%$ \\
0.505 & $0.152\pm0.023$ & $0.210\pm0.047$ & $13\%$ \\
0.51  & $0.089\pm0.002$ & $0.071\pm0.003$ & $13\%$ \\
0.52  & $0.057\pm0.016$ & $0.106\pm0.022$ & $9\%$ \\
0.53  & $0.027\pm0.004$ & $0.117\pm0.036$ & $20\%$ \\
0.56  & $0.054\pm0.010$ & $0.059\pm0.020$ & $12\%$ \\
0.60  & $0.029\pm0.021$ & $0.077\pm0.017$ & $10\%$ \\
0.65  & $0.026\pm0.006$ & $0.042\pm0.012$ & $5\%$ \\
0.70  & $0.024\pm0.006$ & $0.027\pm0.003$ & $4\%$ \\
\end{tabular}
\end{ruledtabular}
\end{table}

\begin{table}[t]
\caption{Distance dependence of the dimer share $f_0$ in the commensurate dimerized regime ($\chi=256$). $d=20$, $60$: sublattice average ($l=2$ / $l=3$).
Largest $d$: share at the largest distance computed ($\chi=192$ agrees within $1.5\times10^{-3}$), for $l=2$ by block position (strong--strong / strong--weak /
weak--weak bonds), for $l=3$ the range over positions. In all cases $f_0$ is still increasing at the largest distance.}
\label{tab:chdist}
\begin{ruledtabular}
\begin{tabular}{lccccc}
$\Jt$ & $\xi$ & $d=20$ & $d=60$ & largest $d$, $l=2$ & largest $d$, $l=3$ \\
\hline
0.32 & -- & 0.680 / 0.667 & 0.808 / 0.800 & -- & -- \\
0.34 & -- & 0.704 / 0.695 & 0.822 / 0.819 & -- & -- \\
0.36 & -- & 0.672 / 0.676 & 0.781 / 0.789 & -- & -- \\
0.38 & 35.6 & 0.609 / 0.634 & 0.705 / 0.737 & 0.867 / 0.795 / 0.661 ($d=170$--$190$) & 0.795--0.801 ($d=190$) \\
0.40 & 18.9 & 0.533 / 0.583 & 0.610 / 0.670 & 0.801 / 0.687 / 0.466 ($d=90$--$101$) & 0.702--0.709 ($d=110$) \\
0.45 & 5.6  & 0.366 / 0.458 & -- & 0.630 / 0.394 / 0.124 ($d=32$--$38$) & 0.466--0.491 ($d=38$--$40$) \\
\end{tabular}
\end{ruledtabular}
\end{table}

\begin{table}[t]
\caption{Scalar share at the Kosterlitz--Thouless point on periodic rings, extrapolated to $N\to\infty$, for blocks of $l$ sites at four cross ratios.
$x\to0$: linear extrapolation through $x=1/16$ and $1/9$; chord: extrapolation to $x=0$ of a quadratic fit to all rings and distances against the
chord cross ratio ($x\le0.3$, no extrapolation in $N$). Errors in brackets: bounds from Appendix~\ref{app:rings}.}
\label{tab:quarter}
\begin{ruledtabular}
\begin{tabular}{lcccccc}
$l$ & $x=4/9$ & $x=1/4$ & $x=1/9$ & $x=1/16$ & $x\to0$ & chord, $x\to0$ \\
\hline
2 & 0.144019(20) & 0.201137(22) & 0.220806(23) & 0.225436(34) & 0.231388(83) & 0.2282 \\
3 & --           & 0.222672(22) & 0.234280(34) & 0.241320(53) & 0.25037(13)  & 0.2466 \\
4 & 0.209885(21) & 0.229344(30) & 0.240579(48) & 0.24413(11)  & 0.24870(25)  & 0.2479 \\
5 & --           & 0.229610(38) & 0.240010(92) & 0.24410(20)  & 0.24935(48)  & 0.2483 \\
6 & 0.212694(53) & 0.230692(91) & 0.24157(53)  & 0.2451(15)   & 0.2497(36)   & 0.2490 \\
\end{tabular}
\end{ruledtabular}
\end{table}

\subsection{Gapless side and KT point: block mutual information versus CFT}
\label{sec:CFTresults}

\begin{figure*}[t]
\centering
\figfile{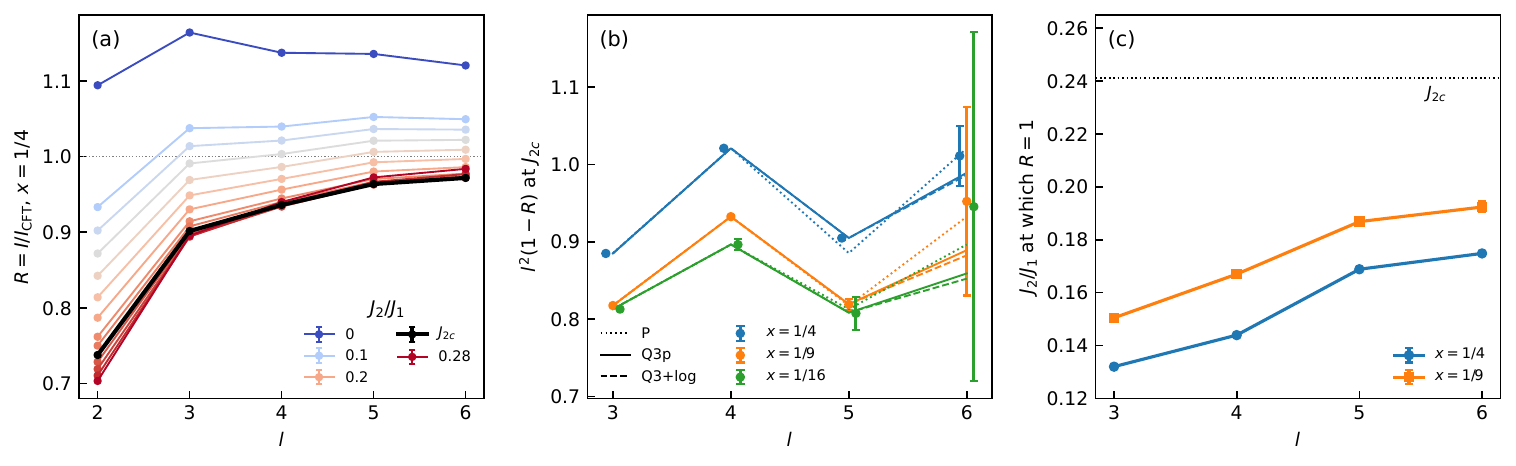}{\textwidth}
\caption{Block mutual information on periodic rings, $N\to\infty$.
(a) $R=I/\Icft$ at $x=1/4$ versus $l$ for $0\le\Jt\le0.28$ (blue to red; thick line: $\Jc$).
(b) $l^2(1-R)$ at $\Jc$ for $x=1/4$, $1/9$ and $1/16$ with the fits P, Q3p and Q3+log of Table~\ref{tab:J2c}.
(c) Coupling at which $R=1$ versus $l$; the dotted line is $\Jc$.}
\label{fig:CFT}
\end{figure*}

Tables~\ref{tab:R} and \ref{tab:R16} list the ratio $R=I/\Icft$ on periodic rings, extrapolated to $N\to\infty$, for $x=1/4$, $1/9$ and $1/16$, $l=3$--$6$, and
Fig.~\ref{fig:CFT}(a) shows it versus $l$. The errors are bounds that contain the bond dimension, the ring size, the placement of the blocks and the
DMRG convergence (Appendix~\ref{app:rings}); they are correlated between neighboring couplings and block lengths, and $\chi^2$ values computed with them
are used only to reject fit forms. For small $\Jt$ the mutual information exceeds the CFT value at all accessible $l$; $R=1$ is crossed at a coupling that
increases with $l$ (Table~\ref{tab:crossing}). Closer to $\Jc$, $R<1$ at these block sizes and approaches $1$ with increasing $l$. Above $\Jc$ the rings
remain free of dimerization (the two placements of the blocks, shifted by one site, differ by at most $8\times10^{-5}$ in the extrapolated $R$), so that
$\Jt>\Jc$ can be used quantitatively. At fixed $l$ and $x$, $R$ has a minimum above $\Jc$ that moves towards $\Jc$ as $l$ grows: for $x=1/4$ it lies
at $\Jt=0.27$, $0.26$, $0.25$, $0.25$ for $l=3$--$6$, and for $x=1/9$ at $0.25$ for $l=3$, $4$ and at the grid point $\Jc$ for $l=5$, $6$. The origin of
this behavior is the structure of the expansion in the coupling discussed below.

\begin{table*}[t]
\caption{Ratio $R=I_l/\Icft$ on periodic rings, extrapolated to $N\to\infty$ at fixed geometry [Eqs.~\eqref{eq:crossratio}, \eqref{eq:Icft}]. Errors in
brackets, in units of the last digits: bounds from bond dimension, ring size, placement and DMRG convergence (Appendix~\ref{app:rings}). Rings: $N=32$,
$48$, $64$, and at $\Jc$ in addition $N=80$, $96$. Values for $x=1/16$ in Table~\ref{tab:R16}.}
\label{tab:R}
\footnotesize
\begin{ruledtabular}
\begin{tabular}{lcccccccc}
 & \multicolumn{4}{c}{$x=1/4$ ($d=l$)} & \multicolumn{4}{c}{$x=1/9$ ($d=2l$)} \\
$\Jt$ & $l=3$ & $l=4$ & $l=5$ & $l=6$ & $l=3$ & $l=4$ & $l=5$ & $l=6$ \\
\hline
0 & 1.16402(12) & 1.13727(20) & 1.13573(33) & 1.12052(87) & 1.22282(41) & 1.19906(63) & 1.18928(82) & 1.1724(26) \\
0.10 & 1.03761(15) & 1.03974(18) & 1.05243(22) & 1.04945(60) & 1.07117(43) & 1.07751(48) & 1.08601(59) & 1.0822(17) \\
0.12 & 1.01385(14) & 1.02128(17) & 1.03649(15) & 1.03567(62) & 1.04224(43) & 1.05369(47) & 1.06538(47) & 1.0639(16) \\
0.14 & 0.99095(12) & 1.00341(13) & 1.02102(14) & 1.02217(57) & 1.01420(44) & 1.03025(44) & 1.04495(42) & 1.0455(14) \\
0.16 & 0.969143(67) & 0.98640(11) & 1.00623(13) & 1.00917(48) & 0.98744(42) & 1.00762(40) & 1.02495(47) & 1.0273(11) \\
0.18 & 0.948776(70) & 0.97056(12) & 0.99249(15) & 0.99695(38) & 0.96253(38) & 0.98638(37) & 1.00610(69) & 1.00989(96) \\
0.20 & 0.930333(48) & 0.956409(98) & 0.98029(17) & 0.98618(57) & 0.94025(20) & 0.96735(43) & 0.98915(58) & 0.9944(18) \\
0.22 & 0.914504(61) & 0.94471(14) & 0.97043(21) & 0.97755(55) & 0.92201(27) & 0.95188(45) & 0.97558(56) & 0.9819(18) \\
0.23 & 0.907865(57) & 0.94012(13) & 0.96675(20) & 0.97447(56) & 0.91499(24) & 0.94621(44) & 0.97083(58) & 0.9775(17) \\
$\Jc$ & 0.901675(37) & 0.936186(66) & 0.963806(88) & 0.9719(11) & 0.909165(96) & 0.94172(13) & 0.96724(30) & 0.9735(34) \\
0.25 & 0.897978(62) & 0.93448(16) & 0.96306(26) & 0.97187(50) & 0.90701(34) & 0.94105(59) & 0.96753(79) & 0.9748(13) \\
0.26 & 0.895194(45) & 0.93401(13) & 0.96370(25) & 0.97317(52) & 0.90709(29) & 0.94283(66) & 0.97047(93) & 0.9781(11) \\
0.27 & 0.894280(44) & 0.93566(12) & 0.96665(22) & 0.97696(42) & 0.91090(24) & 0.94879(60) & 0.97795(91) & 0.98639(65) \\
0.28 & 0.895655(59) & 0.93999(14) & 0.97255(23) & 0.98387(51) & 0.91956(27) & 0.96014(54) & 0.99121(86) & 1.0008(10) \\
\end{tabular}
\end{ruledtabular}
\end{table*}

\begin{table}[t]
\caption{As Table~\ref{tab:R}, for $x=1/16$ ($d=3l$), computed for $\Jt\ge0.20$.}
\label{tab:R16}
\begin{ruledtabular}
\begin{tabular}{lcccc}
$\Jt$ & $l=3$ & $l=4$ & $l=5$ & $l=6$ \\
\hline
0.20 & 0.94693(54) & 0.97450(64) & 0.9952(12) & 1.0002(38) \\
0.22 & 0.92474(51) & 0.95578(63) & 0.97825(54) & 0.9842(20) \\
0.23 & 0.91644(52) & 0.94910(65) & 0.97235(56) & 0.9788(18) \\
$\Jc$ & 0.90968(15) & 0.94397(42) & 0.96770(84) & 0.9737(63) \\
0.25 & 0.90805(65) & 0.94400(92) & 0.9690(11) & 0.9760(16) \\
0.26 & 0.90967(73) & 0.9476(11) & 0.9737(14) & 0.9811(13) \\
0.27 & 0.91682(70) & 0.9574(11) & 0.9849(12) & 0.9934(13) \\
0.28 & 0.93108(62) & 0.9751(11) & 1.0049(18) & 1.0148(32) \\
\end{tabular}
\end{ruledtabular}
\end{table}

\paragraph*{At the KT point: power law and logarithmic term.}
Table~\ref{tab:J2c} collects fits of $1-R(l)$ at $\Jc$ for $l=3$--$6$ with seven forms that combine an $l^{-2}$ or $l^{-p}$ term, an even--odd term and a
logarithm $B/\ln l$. A form is called acceptable if its $\chi^2$ probability exceeds $1\%$. Only the two most rigid forms are rejected, at $x=1/4$ and
for Q3 also at $1/9$. The acceptable forms with a free exponent give $p=1.957$--$2.028$ ($x=1/4$), $1.996$--$2.068$ ($x=1/9$) and $2.012$--$2.066$
($x=1/16$). The forms with a logarithm give amplitudes that are individually significant at $x=1/4$ but of opposite sign, $B=+0.0027(3)$ and
$-0.0017(3)$; the amplitude is thus determined by the choice of form, not by the data, and we quote bounds. Taking the largest $|B|+2\sigma$ over the
forms with a logarithm gives $|B|\le0.0033$, $0.0060$ and $0.0099$ for $x=1/4$, $1/9$ and $1/16$. At $\Jt=0$ the same forms give $B=-0.260$ and $-0.264$
($x=1/4$) and $-0.370$ and $-0.372$ ($x=1/9$); there the forms with a fixed $l^{-2}$ term and no logarithm (P, Q3, Qq) are rejected
($\chi^2\ge7\times10^3$) and a free power law gives $p\simeq0.31$--$0.37$. The logarithmic amplitude at $\Jc$ is therefore suppressed by a factor of at least $79$ ($x=1/4$) and $62$ ($x=1/9$) relative to the
Heisenberg chain, as expected from the vanishing of the marginal coupling. With the rings up to $N=64$ only, the exponents are $1.975$--$2.032$,
$2.022$--$2.077$ and $2.053$--$2.086$ and the suppression factors $75$ and $42$. The even--odd difference of $l^2(1-R)$, $0.885$, $1.021$, $0.905$, $1.012$
for $l=3$--$6$ at $x=1/4$, reflects the parity corrections of the von Neumann
quantity~\cite{Laflorencie2006,Calabrese2010parity,CardyCalabrese2010}. The Rényi-2 mutual information, in contrast, is dominated by strong parity
oscillations at these block sizes (Appendix~\ref{app:renyi}).

\begin{table*}[t]
\caption{Fits of $1-R(l)$ at $\Jc$ for $l=3$--$6$ (rings, $N\to\infty$) at three cross ratios $x$. $\nu$: number of degrees of freedom; the free
parameter is the exponent $q$ or $p$, or the amplitude $B$ of the logarithmic term. Forms with fit probability below $1\%$ are marked $^*$.}
\label{tab:J2c}
\begin{ruledtabular}
\begin{tabular}{llcccccccc}
 & & & & \multicolumn{2}{c}{$x=1/4$} & \multicolumn{2}{c}{$x=1/9$} & \multicolumn{2}{c}{$x=1/16$} \\
\cline{5-6}\cline{7-8}\cline{9-10}
form & fit function & $\nu$ & free & $\chi^2$ & value & $\chi^2$ & value & $\chi^2$ & value \\
\hline
P      & $A\,l^{-2}[1+a(-1)^l]$        & 2 & --  & $80.1^*$ & --           & $0.063$  & --            & $0.11$  & -- \\
Q3     & $A\,l^{-2}+C(-1)^l\,l^{-3}$   & 2 & --  & $25.3^*$ & --           & $11.1^*$ & --            & $1.43$  & -- \\
\colrule
Qq     & $A\,l^{-2}+C(-1)^l\,l^{-q}$   & 1 & $q$ & $0.010$  & $2.61(7)$    & $0.029$  & $2.05(26)$    & $0.039$ & $1.7(9)$ \\
Q3p    & $A\,l^{-p}+C(-1)^l\,l^{-p-1}$ & 1 & $p$ & $0.32$   & $2.028(6)$   & $0.27$   & $2.068(21)$   & $0.15$  & $2.07(6)$ \\
Pp     & $A\,l^{-p}[1+a(-1)^l]$        & 1 & $p$ & $0.53$   & $1.957(5)$   & $0.023$  & $1.996(18)$   & $0.057$ & $2.01(5)$ \\
\colrule
P+log  & P $+\,B/\ln l$                & 1 & $B$ & $0.73$   & $0.0027(3)$  & $0.022$  & $0.0002(10)$  & $0.060$ & $-0.0007(28)$ \\
Q3+log & Q3 $+\,B/\ln l$               & 1 & $B$ & $0.42$   & $-0.0017(3)$ & $0.33$   & $-0.0037(11)$ & $0.17$  & $-0.0036(32)$ \\
\end{tabular}
\end{ruledtabular}
\end{table*}

\paragraph*{The crossing point.}
For $\Jt<\Jc$ the approach to the CFT value is much slower: at $\Jt=0$, $R(x=1/4)=1.164$, $1.137$, $1.136$, $1.121$ for $l=3$--$6$. The value of $\Jt$ at
which $R=1$ for a given $l$ (Table~\ref{tab:crossing}, Fig.~\ref{fig:CFT}(c)) drifts upwards with $l$: from $0.132$ to $0.175$ for $x=1/4$ and from $0.151$
to $0.192$ for $x=1/9$ as $l$ grows from $3$ to $6$. Since the $l^{-2}$ correction vanishes faster than the marginal correction, the crossing point
should converge to $\Jc$; with four block sizes we can only state that it moves monotonically towards $\Jc$.

\begin{table}[t]
\caption{Coupling $\Jt$ at which $I=\Icft$ for given $l$ (rings; cubic interpolation between the computed couplings; range from $R\pm\sigma$ in brackets).}
\label{tab:crossing}
\begin{ruledtabular}
\begin{tabular}{lcc}
$l$ & $x=1/4$ & $x=1/9$ \\
\hline
3 & 0.1320 [0.1319, 0.1321] & 0.1505 [0.1501, 0.1508] \\
4 & 0.1439 [0.1438, 0.1441] & 0.1670 [0.1666, 0.1673] \\
5 & 0.1688 [0.1686, 0.1690] & 0.1869 [0.1861, 0.1876] \\
6 & 0.1748 [0.1742, 0.1755] & 0.1924 [0.1906, 0.1945] \\
\end{tabular}
\end{ruledtabular}
\end{table}

\begin{figure*}[t]
\centering
\figfile{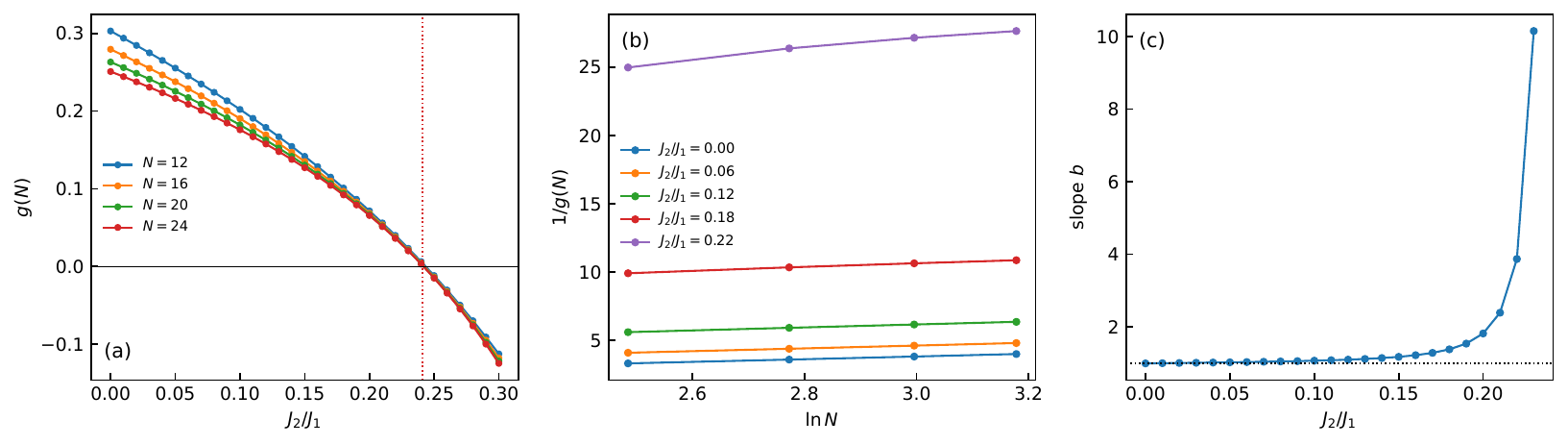}{\textwidth}
\caption{Level spectroscopy on periodic rings ($N=12$--$24$, exact diagonalization). (a) Marginal coupling $g(N)$ of Eq.~\eqref{eq:gN}; its zero
approaches $\Jc$ (dotted line). (b) $1/g$ versus $\ln N$ for selected $\Jt$: parallel lines indicate a single marginal coupling. (c) Fitted slope $b$
versus $\Jt$; it is constant ($\approx1.0$) for $\Jt\le0.12$ and rises close to $\Jc$, where $g$ is small.}
\label{fig:spectroscopy}
\end{figure*}

\paragraph*{The marginal coupling measured by level spectroscopy.}
Figure~\ref{fig:spectroscopy} and Table~\ref{tab:coupling} (Appendix~\ref{app:coupling}) give $g(N)$ from Eq.~\eqref{eq:gN}. Its zero gives
$\Jc(N)=0.24345$, $0.24244$, $0.24196$, $0.24171$ for $N=12$--$24$, converging to $0.241167$~\cite{OkamotoNomura1992}. Plotted as $1/g$ versus $\ln N$,
the data for different $\Jt$ form parallel lines: the fitted slope is $b=0.99$ at $\Jt=0$ and $b=0.99$--$1.09$ for all $\Jt\le0.12$, with median $1.026$,
as required for a single marginal coupling [Eq.~\eqref{eq:running}]. In the normalization $\Delta_{\rm t}=\tfrac12-\tfrac g4$, $\Delta_{\rm s}=\tfrac12+\tfrac{3g}4$
used here, the logarithmic corrections of the Heisenberg chain correspond to $g\simeq1/\ln L$~\cite{Affleck1989}, i.e.\ $b=1$, in agreement with the
measured slope. Closer to $\Jc$ the apparent slope rises (to $3.9$ at $\Jt=0.22$), because $g$ becomes small and higher-order and irrelevant corrections
dominate $1/g$; we therefore obtain $g_0$ with the plateau value $b=1.026$ and extrapolate it linearly in $1/N^2$. The extrapolated $g_0$ decreases from
$1.40$ at $\Jt=0$ to $0.0822$ at $0.20$ and $0.0204$ at $0.23$, passes through zero at $\Jt=0.24117$, and reaches $-0.0619$ at $0.28$; it is a concave
function of $\Jc-\Jt$, with slope $dg_0/d\Jt=-1.771$ at $\Jc$. Using $g_0$ of the largest ring alone ($N=24$) instead changes the slope at $\Jc$ by
$0.4\%$ and moves the zero to $0.24175$.

\begin{figure*}[t]
\centering
\figfile{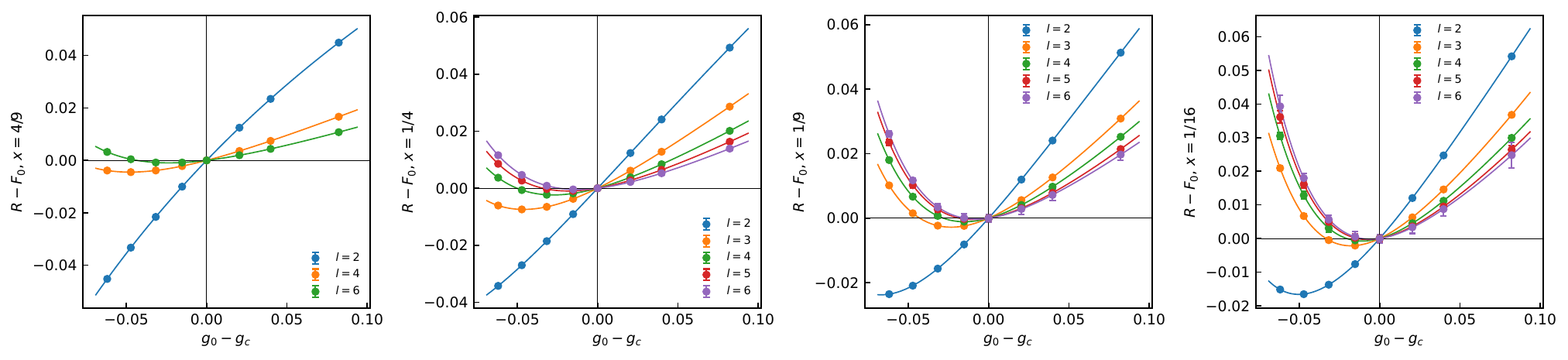}{\textwidth}\\[4pt]
\figfile{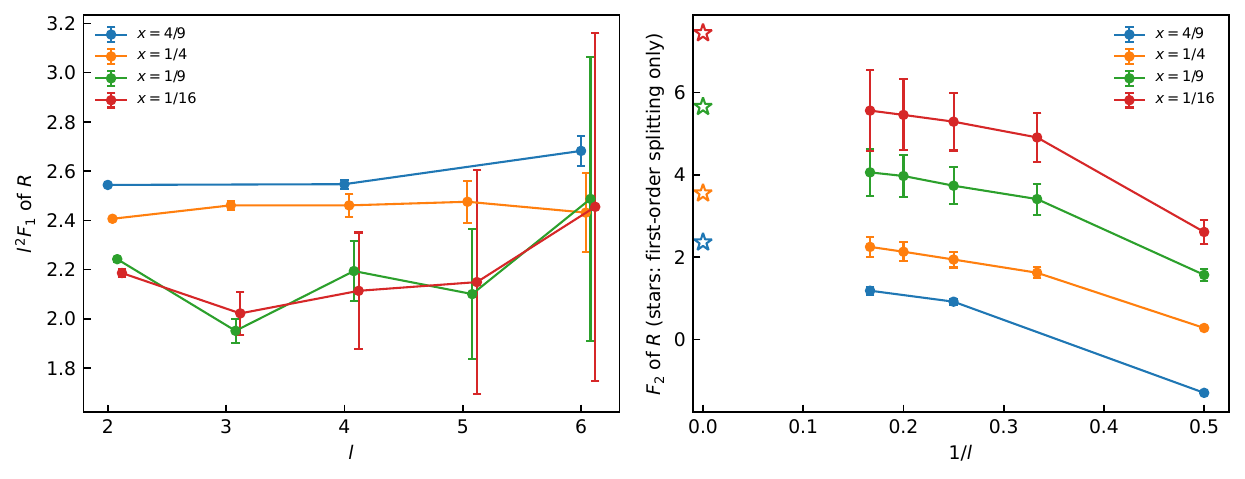}{0.8\textwidth}
\caption{Expansion in the marginal coupling at the KT point (rings, $N\to\infty$).
Top: $R-F_0$ versus $g_0-g_c$ at $x=4/9$, $1/4$, $1/9$ and $1/16$ for the couplings $0.20\le\Jt\le0.28$; lines: fitted polynomials of the central
variant. As $l$ grows, the minima of the fitted curves move towards $g_0=0$.
Bottom left: $l^2F_1$ versus $l$, constant for a pure $l^{-2}$ law. Bottom right: $F_2$ versus $1/l$; stars at $1/l=0$: $F_2^{\rm split}$ from the
first-order splitting alone (not a full prediction).}
\label{fig:coeff}
\end{figure*}

\paragraph*{First and second order in the coupling.}
We determine the coefficients of Eq.~\eqref{eq:Fexp} from the eight couplings $0.20\le\Jt\le0.28$, three below and four above $\Jc$. For every ring
size, bond dimension, geometry and placement, $R$ is fitted by a polynomial of order $3$, $4$ or $5$ in $\Jc-\Jt$, the polynomial is converted to a
series in $g_0$ with the slope and curvature of $g_0(\Jt)$ at $\Jc$, and the coefficients are then extrapolated to $N\to\infty$ (Appendix~\ref{app:rings}).
Because the fits are made at fixed ring size and bond dimension, errors common to all couplings drop out of $F_1$ and $F_2$. A polynomial of order $4$
over the full window describes the extrapolated values of all geometries ($\chi^2$ probability $0.87$--$1$) and is used as the central variant; the
spread over windows ($0.20$ or $0.22\le\Jt\le0.28$) and orders is the model error. A polynomial of low order in $g_0$ itself does not describe the
data at this precision, because the relation between $\Jt$ and $g_0$ is curved, and it overestimates $F_2$ by $20$--$40\%$. Table~\ref{tab:coeff} and
Fig.~\ref{fig:coeff} give the results.

The first-order coefficient decays as $l^{-2}$. The product $l^2F_1$ is constant within its errors, e.g.\ $2.406(4)$, $2.461(19)$, $2.461(46)$,
$2.476(85)$ and $2.43(16)$ for $l=2$--$6$ at $x=1/4$. For $l\ge3$ the one-parameter form $C/l^2$ is accepted at all cross ratios, with $C=2.461(17)$,
$1.99(4)$ and $2.04(8)$ for $x=1/4$, $1/9$ and $1/16$ (probabilities $0.996$, $0.23$, $0.91$; $C=2.54$ at $x=4/9$ from $l=2$, $4$, $6$); a free exponent
gives $p=1.999(46)$, $1.70(15)$ and $1.82(26)$. Adding a constant, $a+C/l^2$, gives $a=0.000(4)$, $0.020(11)$ and $0.013(18)$, i.e.\ $|a|<0.008$ at
$x=1/4$ and $|a|\lesssim0.04$ at the smaller cross ratios. When $l=2$ is included, a small $l^{-3}$ term is needed at $x=1/4$ ($C/l^2+D/l^3$ with $D=-0.28(9)$)
and no simple form is accepted at $x=1/9$, where the constant becomes $a=-0.022(6)$; we therefore quote the bounds from $l\ge3$. The form $a+b/l$, which
mimics $l^{-2}$ over a short range with a spurious negative constant, is rejected at $x=1/4$ ($\chi^2=19.5$ for $2$ degrees of freedom, $l\ge3$) and
at all cross ratios when $l=2$ is included. To first order in the coupling, the marginal perturbation thus changes only the amplitude of the $l^{-2}$
correction, $1-R\simeq[A(x)-C(x)\,g_0]\,l^{-2}$ with $A(1/4)\approx0.95$ and $C(1/4)=2.46$; the data give no indication that a first-order term survives for large blocks, in agreement
with statement (i) of Sec.~\ref{sec:smallx}.

For $x\le1/4$ and $l\ge2$ the second-order coefficient is positive and increases with $l$, by decreasing amounts (Table~\ref{tab:coeff}). These data
are consistent with a finite positive limit, but its value depends on the extrapolation form and is not determined uniquely by blocks with $l\le6$: $a+b/l^2$ fitted to $l=2$--$6$ gives $2.55(13)$, $4.52(32)$ and $6.21(52)$ for $x=1/4$, $1/9$ and $1/16$; $a+b/l$ gives $2.89(39)$, $4.75(94)$ and
$6.27(156)$ for $l\ge3$ and $3.57(20)$, $5.81(47)$ and $7.78(80)$ for $l\ge2$. The extrapolated values thus range over $2.4$--$3.6$, $4.3$--$5.8$ and $5.8$--$7.8$, respectively.
The values from the first-order splitting alone, $F_2^{\rm split}=3.55$, $5.66$ and $7.45$ (Sec.~\ref{sec:smallx}), lie at the upper end of these ranges
and follow the same increase with decreasing $x$; since second-order shifts of the dimensions are not included in $F_2^{\rm split}$, this is a
consistency of size and trend, not a test. No term growing like $\ln l$ is visible, as expected when the first-order term has no $l$-independent part, consistent with the bounds on $a$ above.
Changing the definition of $g_0$ (Appendix~\ref{app:coupling}) changes $F_1$ by at most $0.5\%$ and $F_2$ by at most $0.05$; these changes are common to
all geometries and not included in the errors.

At fixed $x$ the ring values themselves have a minimum of $R(\Jt)$ above $\Jc$ (Table~\ref{tab:R}). Within the quadratic approximation in $g_0$,
the positive values of $F_2$ at all accessible $l$ place this minimum at $g_0\simeq-F_1/2F_2<0$, i.e.\ above $\Jc$. For $x=1/4$ this gives
$\Jt\simeq0.265$, $0.255$ and $0.250$ for $l=4$--$6$, close to the minima of the ring values ($0.26$, $0.25$, $0.25$ on a grid of spacing $0.01$); for
$l=3$ it gives $0.296$, outside the computed range and above the minimum of the ring values at $0.27$, so that higher orders in $g_0$ are not negligible
there. Since $F_1$ decreases as $l^{-2}$ while $F_2$ increases, this minimum moves towards the KT point as $l$
grows [Fig.~\ref{fig:coeff}, top]. The vanishing of $F_1$ alone would make the KT point a stationary point of $R(\Jt)$ for large blocks; that this
stationary point is a minimum requires in addition that $F_2$ remains positive as $l\to\infty$, which the data support but cannot establish beyond
$l=6$. Restricted to $\Jt<\Jc$ and to $0.10\le\Jt\le0.22$, the excess over the KT point is dominated by the first-order term for small $l$
and appears nearly linear in $g_0$ with an effective $l$ dependence; data on one side alone cannot separate the two orders.

\begin{table*}[t]
\caption{Coefficients $l^2F_1$ and $F_2$ of $R=F_0+F_1g_0+F_2g_0^2+\dots$ at $\Jc$ [Eq.~\eqref{eq:Fexp}] (rings, $N\to\infty$). Central variant:
order-4 polynomial in $\Jc-\Jt$ for $0.20\le\Jt\le0.28$, converted to the extrapolated coupling $g_0$. Errors include ring extrapolation, convergence
and model choice, but not the common normalization uncertainty of $g_0$ ($0.5\%$ for $F_1$, $\le0.05$ for $F_2$).}
\label{tab:coeff}
\begin{ruledtabular}
\begin{tabular}{lccccc}
 & $l=2$ & $l=3$ & $l=4$ & $l=5$ & $l=6$ \\
\hline
$l^2F_1$, $x=4/9$  & 2.5439(23) & --         & 2.547(18) & --        & 2.682(61) \\
$l^2F_1$, $x=1/4$  & 2.4064(37) & 2.461(19)  & 2.461(46) & 2.476(85) & 2.43(16) \\
$l^2F_1$, $x=1/9$  & 2.2424(98) & 1.951(49)  & 2.19(12)  & 2.10(26)  & 2.49(58) \\
$l^2F_1$, $x=1/16$ & 2.186(18)  & 2.022(87)  & 2.11(24)  & 2.15(45)  & 2.46(71) \\
\hline
$F_2$, $x=4/9$  & $-1.301(24)$ & --       & 0.916(70) & --       & 1.183(99) \\
$F_2$, $x=1/4$  & 0.277(53)    & 1.62(14) & 1.94(19)  & 2.13(23) & 2.25(24) \\
$F_2$, $x=1/9$  & 1.57(14)     & 3.41(38) & 3.74(45)  & 3.97(50) & 4.06(57) \\
$F_2$, $x=1/16$ & 2.61(30)     & 4.91(59) & 5.29(70)  & 5.46(86) & 5.56(98) \\
\end{tabular}
\end{ruledtabular}
\end{table*}

\begin{figure}[t]
\centering
\figfile{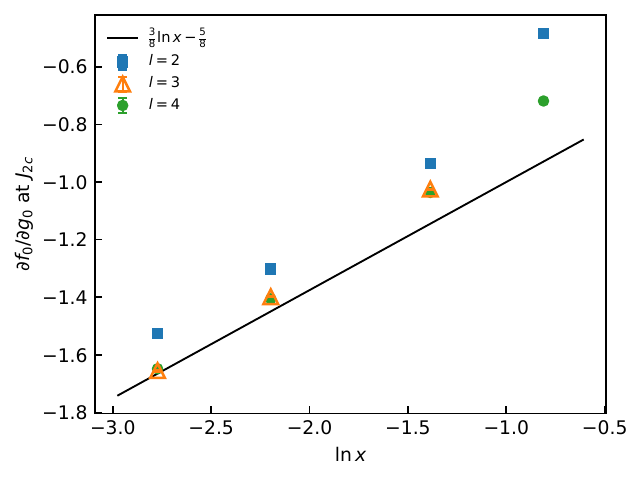}{0.6\columnwidth}
\caption{First-order coefficient of the scalar share at $\Jc$ versus $\ln x$ for $l=2$ (squares), $3$ (open triangles) and $4$ (circles)
(rings, $N\to\infty$). Line: Eq.~\eqref{eq:shareslope}.}
\label{fig:shareslope}
\end{figure}

\paragraph*{The scalar share at first order.}
The same procedure applied to the scalar share gives its first-order coefficient $\partial f_0/\partial g_0$ at $\Jc$ (Table~\ref{tab:shareslope},
Fig.~\ref{fig:shareslope}). Unlike $F_1$ of $R$, it does not vanish for large blocks: it changes by less than $0.012$ between $l=3$ and $4$. Its dependence on
$x$ approaches Eq.~\eqref{eq:shareslope} as $x$ decreases, and much more closely for $l=3$ and $4$ than for $l=2$. At $l=4$ the deviation from $\tfrac38\ln x-\tfrac58$ is $23\%$, $9.5\%$,
$3.0\%$ and $1.0\%$ for $x=4/9$, $1/4$, $1/9$ and $1/16$, decreasing faster than linearly in $x$, as expected for contributions of fields of higher
dimension. A fit $\partial f_0/\partial g_0=A\ln x+B+Cx$ to the four cross ratios at $l=4$ gives $A=0.363(7)$ and $B=-0.670(17)$, compared with $\tfrac38$ and
$-\tfrac58$ ($\chi^2=1.0$ for one degree of freedom); with $A=\tfrac38$ fixed and a quadratic in $x$, $B=-0.632(6)$. At $l=3$ (three cross ratios) the same
fit with $A$ fixed gives $B=-0.639(2)$, whereas at $l=2$ the free fit gives $A=0.181(3)$, far from $\tfrac38$. The data are thus consistent with
Eq.~\eqref{eq:shareslope} at $l=3$ and $4$, at the level of $1.7$ and $2.6$ standard deviations of the bounds for the slope and the intercept, and the
agreement is much better than at $l=2$; channel shares for $l\ge5$ away from $\Jc$ would be needed to establish the limit. The normalization uncertainty
of $g_0$ ($0.5\%$) is not included.

\begin{table}[t]
\caption{First-order coefficient of the scalar share, $\partial f_0/\partial g_0$ at $\Jc$ (rings, $N\to\infty$), compared with
Eq.~\eqref{eq:shareslope}.}
\label{tab:shareslope}
\begin{ruledtabular}
\begin{tabular}{lcccc}
$x$ & $\tfrac38\ln x-\tfrac58$ & $l=2$ & $l=3$ & $l=4$ \\
\hline
$4/9$  & $-0.929$ & $-0.48439(82)$ & --            & $-0.7190(10)$ \\
$1/4$  & $-1.145$ & $-0.93566(81)$ & $-1.0247(11)$ & $-1.0361(14)$ \\
$1/9$  & $-1.449$ & $-1.30158(90)$ & $-1.3979(16)$ & $-1.4054(28)$ \\
$1/16$ & $-1.665$ & $-1.5261(17)$  & $-1.6542(26)$ & $-1.6477(63)$ \\
\end{tabular}
\end{ruledtabular}
\end{table}

\subsection{Discussion}
\label{sec:discussion}

The results above differ in status. Two statements are exact consequences of SU(2) and time-reversal symmetry: the two-site mutual information is
a function of the spin correlator [Eq.~\eqref{eq:fC}, Sec.~\ref{sec:MI}], and for $l=2$ the scalar channel contains only bond-energy correlations, with
no cross-block correlation between chirality and bond energies (Sec.~\ref{sec:channels}). Numerically established are the equality of the leading
triplet and dimer correlation lengths (Sec.~\ref{sec:xi}), the onset of incommensurate real-space decay at the MG point (Sec.~\ref{sec:MG}), the
Lifshitz point (Sec.~\ref{sec:Lifshitz}), the dominance of the dimer channel at the largest distances computed for $0.32\le\Jt\le0.40$, the slope
of the marginal coupling, and the scalar share $1/4$ at $\Jc$ (Secs.~\ref{sec:channelresults} and \ref{sec:CFTresults}). The $l^{-p}$ correction at
$\Jc$, the suppression of the logarithmic term and the $l$ dependence of the first- and second-order coefficients are established only for $l\le6$
(Sec.~\ref{sec:CFTresults}). The small-$x$ results are derived under stated assumptions (Sec.~\ref{sec:smallx}), and the origin of the small scalar
share at $\Jt=0$, the growth of the dimer share with distance and the chirality contribution are interpretations (Sec.~\ref{sec:channelresults}).

\paragraph*{What mutual information measures across the crossover.}
Two-site mutual information is a function of the spin correlator and therefore follows the real-space correlation function, including its nodes. The
channel analysis shows that the additional correlations carried by blocks are primarily dimer-like, and that they do not introduce a new long-distance
scale: spin and dimer correlations are governed by the same two-spinon threshold. In the commensurate dimerized regime up to $\Jt=0.40$ the dimer channel becomes the
dominant contribution at distances of several correlation lengths, which we attribute to a more slowly decaying power-law prefactor. Its suppression,
which begins within the commensurate regime and is nearly complete between the disorder and Lifshitz points, reflects a redistribution of correlation
content between channels, not the appearance of a new correlation length.
Because block mutual information is a real-space quantity, it already follows the incommensurate decay between the disorder point, which marks the onset
of incommensurate real-space decay, and the Lifshitz point, which marks the splitting of the structure-factor peak.

\paragraph*{The MG point.}
Around the MG point, the linear vanishing of $C(2)$ and the collapse of the first-order correlation profile show that the two sides share the same leading
perturbative structure. The distinction between commensurate and incommensurate decay enters only at higher order and in the far tail of the correlations.
The transfer-matrix results are consistent with the disorder point being located at the MG point, but the onset law of $\pi-q$ cannot be determined,
because the modulation length diverges while the correlation length remains short. An analytic continuation of the MG spinon dispersion to complex
momenta would be the natural tool for this question.

\paragraph*{The KT point as a reference for CFT tests.}
At $\Jc$ the marginal coupling vanishes, so the leading finite-block correction is no longer masked by a slowly decaying logarithm. This makes the KT
point the natural place to compare with the two-interval CFT prediction: on dimerization-free rings and for $l\le6$ the correction follows $l^{-p}$ with
$p$ close to $2$, as expected for the leading irrelevant operators, the logarithmic amplitude is suppressed by more than an order of magnitude, and the
deviation from $R=1$, $3\%$ at $l=6$ and decreasing, is consistent with $\mathcal{F}_1\equiv1$ at the self-dual radius. The scalar share at $\Jc$ approaches
$1/4$ at small $x$, a second, independent signature of the SU(2)$_1$ fixed point, and its strong suppression at $\Jt=0$ makes the marginal coupling
visible in the operator content of the mutual information.

\paragraph*{How the marginal coupling enters.}
Data on both sides of $\Jc$ indicate that the coupling does not enter the two-interval mutual information at first order for large blocks: for $l\le6$
the first-order coefficient only modifies the amplitude of the $l^{-2}$ correction, while the second-order coefficient remains positive. Consistently, the
minimum of $R(\Jt)$ moves towards $\Jc$ as $l$ grows, which suggests that for large blocks the KT point becomes a stationary point of $R(\Jt)$, and a
minimum if $F_2$ stays positive. The leading small-$x$ term of the mutual information explains this structure if the coupling enters
only through the scaling dimensions: the shifts of the four dimension-$\tfrac12$ fields add up to zero, so $R$ has no first-order term, whereas the
scalar share, which compares the singlet with the triplet, changes at first order by $\tfrac38\ln x-\tfrac58$. The measured share coefficients approach this value as $x$ decreases, closely for $l=3$ and $4$ and not for $l=2$. Whether the assumption holds, i.e.\ whether there are additional first-order contributions from the branch
points of the replica geometry, is a question for field theory; the lattice results constrain them to be small at the accessible block sizes. With data
below $\Jc$ only, the same results appear as an excess nearly linear in the coupling with an effective $l$ dependence, which does not reveal this
structure.

\paragraph*{Limitations.}
Three classes of limitations remain. (i) Finite-entanglement and finite-size limitations: block sizes are restricted to $l\le6$, so the asymptotic forms of
the corrections are constrained by four or five block sizes; away from $\Jc$ the rings reach $N=64$, and the channel shares are available only for $l\le4$;
the long-distance channel shares in the dimerized phase are still increasing at the largest distances computed. (ii) Analysis-dependent systematics: the
ring errors are bounds that combine bond dimension, ring size, placement and DMRG convergence, and they are correlated between couplings and block
lengths, and the bond-dimension part may be underestimated by up to a factor $1.5$ (Appendix~\ref{app:rings}); the channel shares rely on a
second-order expansion; the coupling $g_0$ relies on its extraction from level spectroscopy, whose normalization
uncertainty ($0.5\%$ for $F_1$) is common to all geometries. (iii) Unresolved asymptotics: the limit of the second-order coefficient for $l\to\infty$
depends on the extrapolation form; the first-order change of the scalar share is consistent with Eq.~\eqref{eq:shareslope} at $l=3$, $4$ but its limit is
not established; the derivation of Sec.~\ref{sec:smallx} rests on assumptions that remain to be justified; quantities above $\Jc$ are restricted to
$\Jt\le0.28$, where the gap exceeds all accessible lengths; and the onset law of the incommensurate wave vector is not accessible.

\section{Conclusions}
\label{sec:conclusions}

Taken together, our results show that in the $J_1$--$J_2$ chain enlarging the subsystems from individual sites to finite blocks enriches the correlation
content resolved by mutual information without generating an independent long-distance scale. For SU(2)-singlet states, symmetry alone reduces the two-site mutual information to
the spin correlator, so any new content must come from blocks. What blocks add is structural rather than a new spatial scale. The dominance of the scalar
contribution at the largest distances computed in the commensurate dimerized regime up to $\Jt=0.40$, where the share is still increasing with
distance, shows that block mutual information resolves substantial bond-energy correlations that are inaccessible
to the two-site spin correlator, while its decay is still governed by the length scale that spin and dimer correlations share.

The commensurate--incommensurate crossover illustrates the same point from a different angle.
The distinction between the disorder and Lifshitz points shows that mutual information follows
the real-space correlation structure rather than defining an independent crossover scale:
the redistribution of its channel content begins within the commensurate regime and is
largely complete between the onset of incommensurate real-space decay at the disorder point and
the splitting of the structure-factor peak at the Lifshitz point. Mutual information is therefore best viewed here not as a probe
of new length scales, but as a way to decompose a known correlation structure into physically
distinct channels.

On the critical side, the vanishing marginal coupling at the KT point provides a controlled lattice benchmark in which the leading finite-block correction
to the two-interval SU(2)$_1$ CFT prediction can be isolated with minimal contamination from logarithmic corrections: on periodic rings with $l\le6$
it follows $l^{-p}$ with $p$ close to $2$, and the scalar share approaches the value $1/4$ of the SO(4)-symmetric fixed point. Following the mutual
information through the KT point, with the marginal coupling measured independently by level spectroscopy, separates the orders of the coupling: the
first-order term only renormalizes the amplitude of the $l^{-2}$ correction, the second-order term stays positive and is consistent with a finite
limit, whose value the present block sizes do not fix; this suggests that for large blocks the KT point becomes a stationary point of the mutual
information relative to its field-theory value. Under the assumptions stated in
Sec.~\ref{sec:smallx}, the splitting of the SU(2)$_1$ multiplet by the marginal coupling accounts for this structure within the leading small-$x$ term
and predicts the first-order change of the scalar share; the data for $l=3$ and $4$ agree with it within a few percent at small $x$, much better than
those for $l=2$. Larger blocks, channel shares beyond $l=4$ away from the KT point, and a field-theoretical treatment of the
first-order corrections on the replica geometry would test whether these features persist in the asymptotic regime.

\begin{acknowledgments}
This research was supported by the Research Grant, Faculty of Science, Prince of Songkla University (contract no. SCITUG65001).
\end{acknowledgments}

\appendix

\section{Two-site mutual information of SU(2) singlets}
\label{app:fC}

Let $|\psi\rangle$ be invariant under global SU(2) rotations. Then $\rho_{ij}$ commutes with $\Sv_i+\Sv_j$ and is diagonal in the
singlet/triplet basis with equal weights on the three triplet states, $\rho_{ij}=\lambda_sP_s+\lambda_tP_t$, $\lambda_s+3\lambda_t=1$. Since
$\Sv_i\cdot\Sv_j=-\tfrac34P_s+\tfrac14P_t$,
\begin{equation}
C=\langle\Sv_i\cdot\Sv_j\rangle=-\tfrac34\lambda_s+\tfrac34\lambda_t=\tfrac14-\lambda_s,
\end{equation}
which gives Eq.~\eqref{eq:werner}. The single-site density matrices are $\mathbb{1}/2$, so $S_i=S_j=\ln2$ and
\begin{equation}
I(i{:}j)=2\ln2+\lambda_s\ln\lambda_s+3\lambda_t\ln\lambda_t .
\end{equation}
Writing $\lambda=\tfrac14+u$ with $u_s=-C$, $u_t=C/3$ and expanding $-\sum\lambda\ln\lambda=\ln4-2\sum u^2+O(u^3)$,
\begin{equation}
I(i{:}j)=2\Big(C^2+3\frac{C^2}{9}\Big)+O(C^3)=\frac{8}{3}C^2+O(C^3).
\end{equation}
The identity requires a singlet: for an $S=\tfrac12$ doublet (odd open chains) $3\langle S^z_iS^z_j\rangle\ne\langle\Sv_i\cdot\Sv_j\rangle$ and $\rho_{ij}$ is not of Werner form.
Numerically, Eq.~\eqref{eq:fC} agrees with the directly computed $I(i{:}j)$ to $10^{-15}$ for all singlet states studied.

\section{Contraction of two-block reduced density matrices}
\label{app:contraction}

Let $B^s_{[n]}$ be the right-canonical tensors and $\Lambda_{[n]}$ the Schmidt values on the bond to the left of site $n$. For $A_1=[s_0,s_0+l)$ define the
open left block
\begin{align}
T^{(0)}_{aa'}&=\big(\Lambda^2_{[s_0]}\big)_{aa'},\\
T^{(k+1)}_{bb';\,\bm\sigma s,\bm\sigma' s'}&=\sum_{aa'}B^{s}_{ab}\,T^{(k)}_{aa';\bm\sigma\bm\sigma'}\,\overline{B^{s'}_{a'b'}},
\end{align}
and propagate it through the gap with $T\to\sum_s B^{s\,\top}TB^{s*}$ (physical indices traced). The open right block of $A_2$ is built from the identity
on the bond to the right of $A_2$, adding sites from the right. Contracting the two over the shared bond gives $\rho_{A_1A_2}$ as a $4^l\times4^l$ matrix.
The right block depends only on the sublattice of $A_2$ and is computed once per sublattice. The left block is propagated once through all gaps $d$, and
$\rho_{A_1A_2}$ is formed only for the requested distances. The cost is $O(d\,\chi^34^l)$ for the propagation and $O(\chi^2 16^l)$ per distance for the final
contraction. The peak memory is $\approx5\times8\chi^24^l+3\times8\cdot16^l$ bytes, i.e.\ $\approx3.1$~GB for $\chi=128$, $l=6$ (measured $0.9$~GB for
$\chi=64$, $l=6$). Rényi-2 entropies are obtained from $\Tr\rho^2=\sum_{ij}|\rho_{ij}|^2$. Validation: block entropies of non-contiguous regions agree with
TeNPy's \texttt{entanglement\_entropy\_segment} to $4\times10^{-15}$; two-point functions agree with TeNPy's \texttt{correlation\_function} to $10^{-12}$.

\section{SU(2) channel decomposition}
\label{app:channeldecomp}

\begin{figure*}[t]
\centering
\figfile{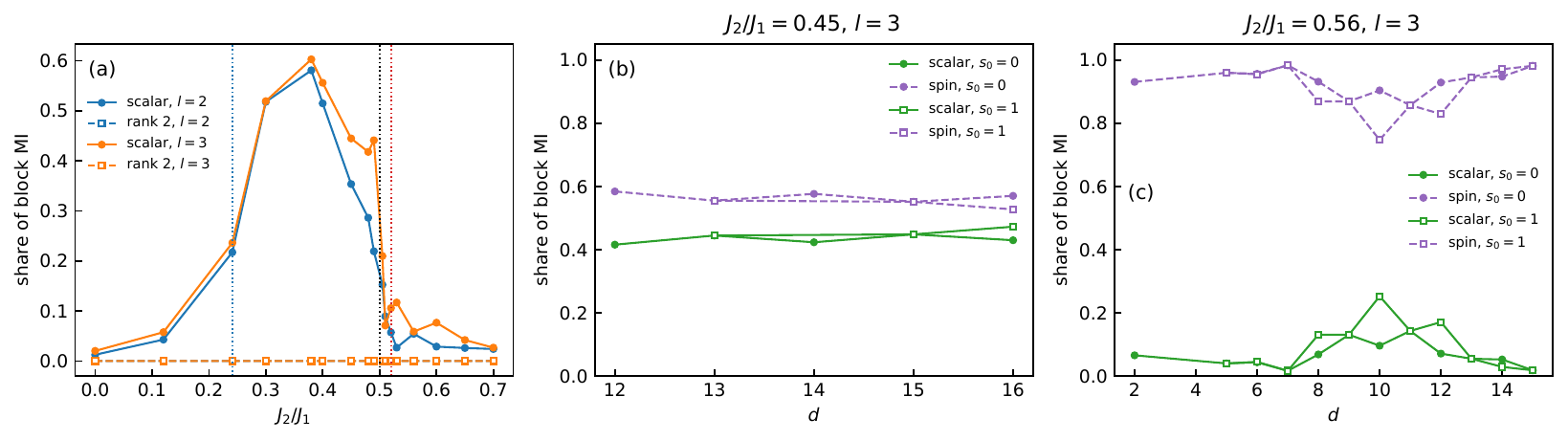}{\textwidth}
\caption{SU(2) channel decomposition of block mutual information ($\chi=128$). (a) Share of the scalar (bond-energy) and rank-2 channels versus $\Jt$ for
$l=2$ and $3$ (10\% acceptance criterion). (b) Scalar and spin shares versus $d$ at $\Jt=0.45$, $l=3$, for both sublattices. (c) The same at $\Jt=0.56$,
where the shares fluctuate with the incommensurate modulation.}
\label{fig:channelsd}
\end{figure*}

For a block of $l$ spins the total-spin operators $S^a_A$ are built in the same basis as $\rho_{A_1A_2}$. Vectorizing operators row-wise,
$(XOY)\mapsto(X\otimes Y^{\top})\,\mathrm{vec}(O)$, the adjoint Casimir acting on the $4^l$-dimensional operator space of a block becomes the matrix
$\mathcal{C}=\Sv_A^2\otimes\mathbb{1}-2\sum_aS^a_A\otimes(S^a_A)^{\top}+\mathbb{1}\otimes(\Sv_A^2)^{\top}$, whose eigenvalues are $k(k+1)$; its
eigenvectors give the projectors $\mathcal{P}_k$. Reshaping $G_{(s_1s_2),(s_1's_2')}$ into $M_{(s_1s_1'),(s_2s_2')}$, the channels are
$M_{kk'}=\mathcal{P}_kM\mathcal{P}_{k'}^{\top}$. For an SU(2)-invariant state only $k=k'$ survives; the largest relative weight of $k\ne k'$ is $2\times10^{-6}$ for the $l\le3$ calculations ($10^{-9}$ for most points) and $4\times10^{-4}$ for the $l\le6$ scans at $\Jc$ and $\Jt=0$. The second-order shares of Eq.~\eqref{eq:channelMI} use the eigen-decomposition of
$\rho_{A_1}\otimes\rho_{A_2}$; eigenvalues are clipped at $10^{-15}$. Checks: for $l=1$ the spin channel carries the full weight; at large $d$,
$\sum_kI_k/I=1.01$ (median); only distances with $|\sum_kI_k/I-1|<0.1$ and $I>10^{-10}$ are used, and the shares quoted are averages over the four
largest valid distances of each sublattice. Figure~\ref{fig:channelsd} shows the distance dependence of the shares at a commensurate and an
incommensurate point.

\paragraph*{Robustness against the acceptance criterion.}
Repeating the analysis with acceptance thresholds of $5\%$, $7.5\%$, $10\%$ and $15\%$, and with either the four largest valid distances or only the
largest one, changes the scalar share by a median of $0.012$ (half-range over the eight variants). The largest spreads occur just below the MG point
($\pm0.105$ at $\Jt=0.49$, $l=2$) and in the incommensurate regime ($\pm0.02$--$0.05$), where only a few distances satisfy the criterion. The shares at
$\Jt=0.38$, $0.603\pm0.010$ ($l=3$) and $0.581\pm0.038$ ($l=2$) for $d\le16$, are robust against the criterion; their increase at larger $d$ is discussed
in Sec.~\ref{sec:channelresults}.

\begin{figure*}[t]
\centering
\figfile{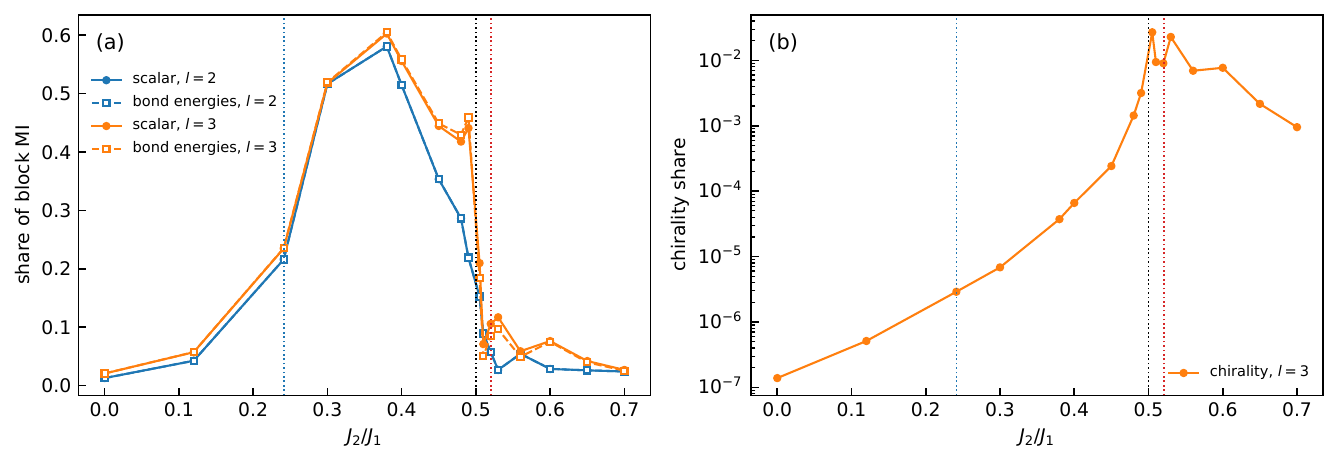}{\textwidth}
\caption{Composition of the scalar channel. (a) Scalar share and the share of the bond energies alone for $l=2$ (identical by construction) and $l=3$.
(b) Share of the scalar chirality alone for $l=3$ (logarithmic scale). Dotted lines mark $\Jc$, the MG point and $\JL$.}
\label{fig:scalarsplit}
\end{figure*}

\paragraph*{Scalar operators and the composition of the scalar channel.}
We verified that the identity and $\Sv_1\cdot\Sv_2$ span the rank-0 space for $l=2$ (dimension 2), and that the identity, the three bond energies and
$\mathcal{K}=\Sv_1\cdot(\Sv_2\times\Sv_3)$ span it for $l=3$ (dimension 5). The chirality is Hermitian and purely imaginary in the real $S^z$ basis, i.e.\ odd
under time reversal, and orthogonal to the bond energies; the projector onto the rank-0 space equals the sum of the projectors onto the bond-energy and
chirality subspaces to $10^{-10}$. Cross terms between bond energies on $A_1$ and chirality on $A_2$ vanish by time reversal (relative weight
$1.5\times10^{-17}$). In the Kubo--Mori metric the bond-energy and chirality parts of $G_0$ are not orthogonal, so we report each alone and the
interference. For $l=3$ the chirality alone carries $10^{-7}$ of the mutual information at $\Jt=0$, rising by five orders of magnitude to $2.7\times10^{-2}$
at $\Jt=0.505$ and decreasing to $10^{-3}$ at $0.70$ [Fig.~\ref{fig:scalarsplit}]; the interference is at most $2\%$ of the mutual information.

\section{Transfer-matrix channels, extrapolation and mode switching}
\label{app:channels}

\paragraph*{Channel weights.}
For the map~\eqref{eq:TM} restricted to $\Delta S^z=0$ we compute right and left eigenvectors $v_k$, $u_k$ and, for $O=S^z$ and $O=h$, the
boundary vectors $\ell_O$ (left environment with $O$ inserted in the first unit cell) and $r_O$ (identity with $O$ inserted). The weight
$w_k(O)$ is the contribution of mode $k$ to $\langle O_0O_{2n}\rangle$. After normalizing by $\sum_k|w_k(O)|$, a mode is labeled triplet if
$|w_k(S^z)|\ge|w_k(h)|$, singlet otherwise, and dark if both normalized weights are below $10^{-7}$. Modes with negligible weight for both
operators (dark modes) occur in both sectors. The correlation lengths of the main text use the symmetry-based identification of Sec.~\ref{sec:TM}, which
is independent of these weights; for 80 of 81 states of the dimerized-phase grid it agrees with the weight-based triplet length to better than $0.5\%$.
The exceptions are $\Jt=0.495$, where nearly degenerate modes made the weight-based label ambiguous, and $0.515\le\Jt\le0.53$, where the leading mode of
the $\Delta S^z=\pm1$ sector has a small weight in $\langle\Sv_0\cdot\Sv_r\rangle$ and the leading mode with appreciable weight is $2$--$3\%$ shorter. A
classification of all modes by their weights for the spin, the dimerization, the vector chirality and the scalar chirality shows that this leading
mode belongs to the vector chirality, which is a rank-1 operator and therefore appears in the $\Delta S^z=\pm1$ sector; for $\Jt\ge0.51$ this is the
case in 47 of 48 states at $\chi=256$. The transfer-matrix correlation length of the $\Delta S^z=1$ sector agrees with TeNPy's
\texttt{correlation\_length} to $10^{-6}$.

\paragraph*{Extrapolation.}
For each $\Jt$ and channel, $m_1=1/\xi(\chi)$ is extrapolated linearly in $\Delta m(\chi)$ to $\Delta m=0$ through the three largest $\chi$
when $\Delta m$ decreases monotonically; otherwise the largest-$\chi$ value is used with the spread of the last two $\chi$ as error. The extrapolation is
essential for $\Jt\lesssim0.45$ and negligible for $\Jt\gtrsim0.48$.

\paragraph*{Mode switching near the MG point.}
For $|\Jt-\tfrac12|\lesssim3\times10^{-4}$ the identity of the leading triplet mode fluctuates from point to point (real modes, complex pairs, and
modes with very small weights, including complex pairs below $\tfrac12$), and a discrete switch of the leading pair at $\Jt-\tfrac12\simeq0.0018$ produces
a jump of $\pi-q$ from $0.019\pi$ to $0.028\pi$ at every $\chi$. The three-parameter fits $A(\Jt-\Jd)^\beta$ in windows smaller than $0.01$ therefore
hit their bounds and are discarded; the local exponent $d\ln(\pi-q)/d\ln(\Jt-\tfrac12)$ wanders between $0.5$ and $0.85$.

\begin{figure*}[t]
\centering
\figfile{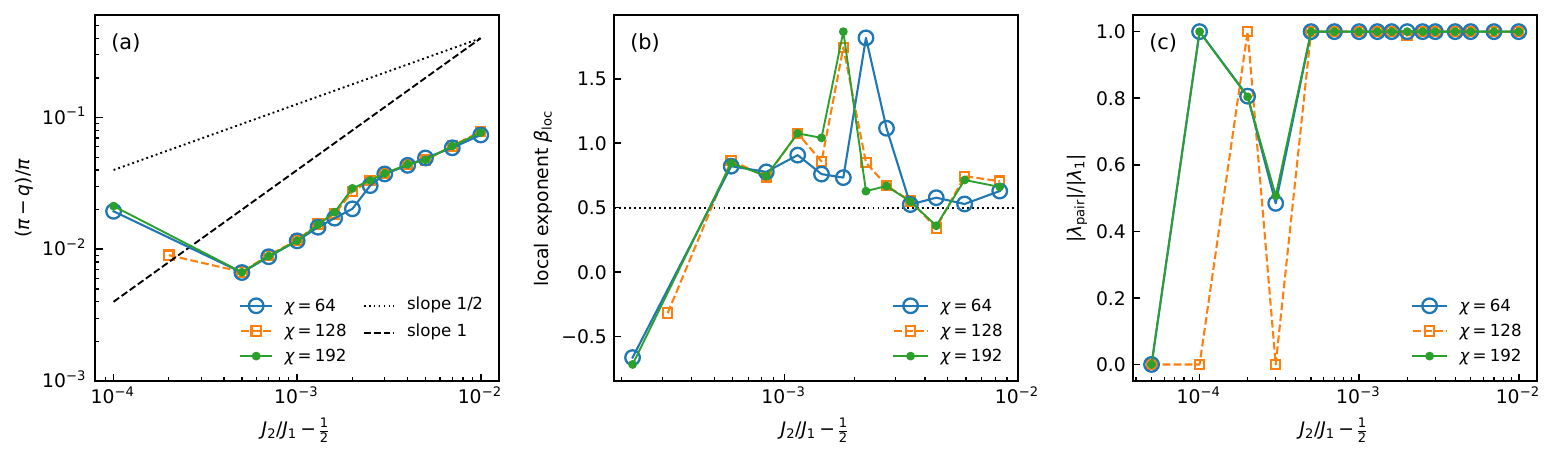}{\textwidth}
\caption{Onset of the incommensurate wave vector near the MG point ($\chi=64$, 128, 192): (a) $(\pi-q)/\pi$ versus $\Jt-\tfrac12$, with guide lines of slope $\tfrac12$ and 1; (b) local exponent; (c) weight ratio $|\lambda_{\rm pair}|/|\lambda_1|$. The peak in (b) near $\Jt-\tfrac12\simeq2\times10^{-3}$ and the scatter below $3\times10^{-4}$ are mode switches.}
\label{fig:A1}
\end{figure*}

\section{Convergence, state selection and the large-$\chi$ estimate}
\label{app:convergence}

\paragraph*{Observable-based convergence.}
At $\Jt=0$ and $\chi=32$, the mutual-information probe $I_3(3)$ changes by $2\times10^{-6}$ over 300 sweeps while $\Delta S$ stays at $4\times10^{-5}$.
A criterion based on $\Delta S<10^{-5}$ therefore declares good states unconverged and, conversely, does not detect metastable states. Such a state
occurred for $\Jt=0$, $\chi=128$ in an early run: its dimerization was $0.023$ and its energy per site $6.3\times10^{-7}$ above that of the state found
later (dimerization $0.0009$); the mutual information differed by up to $3.7\times10^{-3}$. The state-selection procedure of Sec.~\ref{sec:methods}
removes such states.

\paragraph*{Finite-$\chi$ dimerization.}
On the gapless side the best finite-$\chi$ iMPS with a two-site unit cell is sometimes weakly dimerized (Table~\ref{tab:dimer}); at $\Jc$ the
dimerization is $0.02$--$0.03$ for $\chi=96$--$192$. A later run at $\chi=192$ converged to an undimerized state ($D=3\times10^{-4}$) whose energy per site
is lower by only $3.6\times10^{-9}$, below our equivalence threshold; the two states differ in $R$ by at most $4\times10^{-4}$. The dimerization is thus an
almost degenerate direction of the finite-$\chi$ variational space, selected by the history of the run (Appendix~\ref{app:alternation}).

\begin{table}[t]
\caption{Dimer order $D$ of the selected iMPS on the gapless side. At $\Jc$ and $\chi=192$ an undimerized state ($D=3\times10^{-4}$) with an
energy lower by $3.6\times10^{-9}$ per site was found later (see text).}
\label{tab:dimer}
\begin{ruledtabular}
\begin{tabular}{lcccc}
$\Jt$ & $\chi=96$ & $128$ & $160$ & $192$ \\
\hline
0.00  & 0.0008 & 0.0009 & 0.0010 & 0.0026 \\
0.06  & 0.0008 & 0.0008 & 0.0052 & 0.0039 \\
0.10  & 0.0013 & 0.0024 & 0.0017 & 0.0001 \\
0.12  & 0.0033 & 0.0161 & 0.0077 & 0.0036 \\
0.14  & 0.0059 & 0.0059 & 0.0041 & 0.0035 \\
0.16  & 0.0221 & 0.0039 & 0.0223 & 0.0006 \\
0.18  & 0.0134 & 0.0043 & 0.0084 & 0.0119 \\
0.20  & 0.0144 & 0.0178 & 0.0094 & 0.0057 \\
0.22  & 0.0155 & 0.0177 & 0.0007 & 0.0107 \\
$\Jc$ & 0.0239 & 0.0212 & 0.0280 & 0.0269 \\
\end{tabular}
\end{ruledtabular}
\end{table}

\begin{figure*}[t]
\centering
\figfile{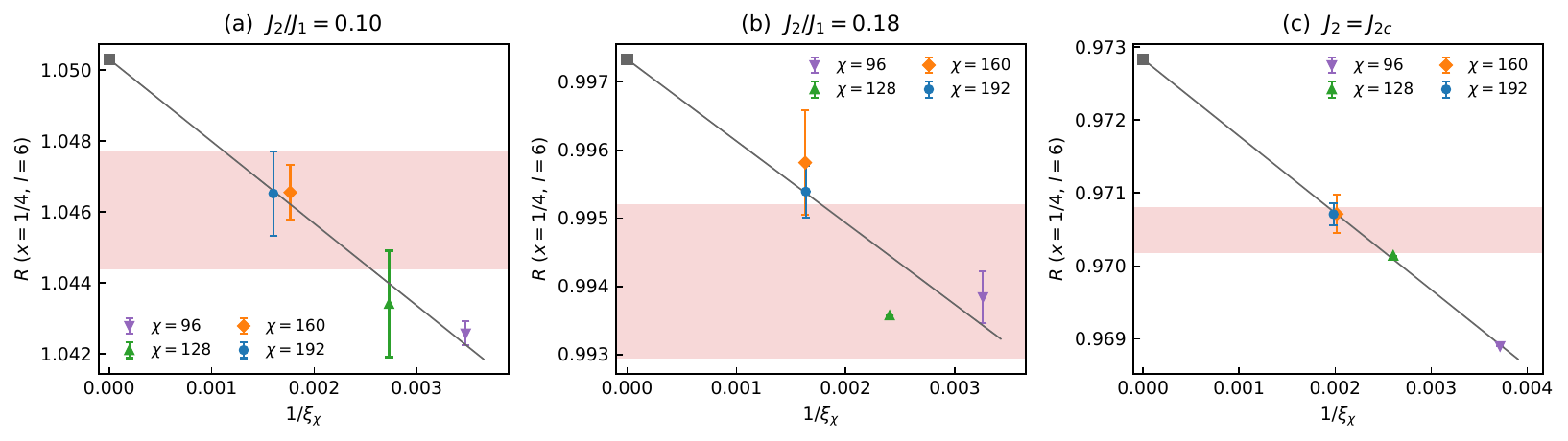}{\textwidth}
\caption{$R$ versus $1/\xichi$ at $x=1/4$, $l=6$ for $\Jt=0.10$, $0.18$ and $\Jc$ (markers as in (a)). Shaded band: large-$\chi$ estimate used in the text; gray line and square: linear fit in $1/\xichi$ and its extrapolation (diagnostic only).}
\label{fig:S1}
\end{figure*}

\paragraph*{Large-$\chi$ estimate versus extrapolation.}
Table~\ref{tab:Rchi} shows $R$ for $x=1/4$ at four bond dimensions. Above $\chi=128$ the values are flat within $\sim10^{-3}$ and at $\Jc$ identical to
four digits for $\chi=160$ and $192$. The largest correlation length $\xichi$ of the iMPS is not monotonic in $\chi$ (e.g.\ $299$, $299$, $573$, $647$ at
$\Jt=0.12$), so a linear extrapolation in $1/\xichi$ amplifies noise (shifts up to $+0.013$). The plateau values are the $\sigma$-weighted means over $\chi\ge128$ with an
error given by half the range over those $\chi$ combined with the statistical error (Fig.~\ref{fig:S1}). Compared with the periodic rings extrapolated to
$N\to\infty$ (Table~\ref{tab:R}), the plateau values lie systematically low (Sec.~\ref{sec:methods}); the extrapolation in $1/\xichi$ removes most of this
bias, at $\Jc$ to within $0.0016$ for all geometries, but with errors of $0.0002$--$0.001$ at $\Jc$ and $\approx0.026$ at $\Jt=0$. Of the 30 iMPS on the
gapless side with $\chi\ge128$, 12 were still drifting at the end of the runs, which is a further reason to base the quantitative analysis on
rings.

\begin{table}[t]
\caption{$R=I/\Icft$ of the infinite chain (iMPS) at $x=1/4$ versus bond dimension; ring values for comparison in Table~\ref{tab:R}.}
\label{tab:Rchi}
\begin{ruledtabular}
\begin{tabular}{llcccc}
$\Jt$ & $l$ & $\chi=96$ & $128$ & $160$ & $192$ \\
\hline
0.00  & 4 & 1.1298 & 1.1332 & 1.1330 & 1.1343 \\
0.00  & 6 & 1.1109 & 1.1156 & 1.1153 & 1.1171 \\
0.12  & 4 & 1.0166 & 1.0182 & 1.0199 & 1.0193 \\
0.12  & 6 & 1.0291 & 1.0314 & 1.0339 & 1.0330 \\
0.18  & 4 & 0.9686 & 0.9682 & 0.9697 & 0.9694 \\
0.18  & 6 & 0.9938 & 0.9936 & 0.9958 & 0.9954 \\
$\Jc$ & 4 & 0.9342 & 0.9349 & 0.9352 & 0.9352 \\
$\Jc$ & 6 & 0.9689 & 0.9701 & 0.9707 & 0.9707 \\
\end{tabular}
\end{ruledtabular}
\end{table}

\section{Periodic rings: extrapolation, errors and coefficients}
\label{app:rings}

\paragraph*{Extrapolation in the ring size.}
For every coupling and geometry the value at $N\to\infty$ is the straight line in $1/N^2$ through the largest ring $N_2$ and the largest ring
$N_1\le0.8\,N_2$, each at its largest bond dimension. The condition on $N_1$ keeps the lever arm $\lambda=N_1^2/(N_2^2-N_1^2)$, with
value $=(1+\lambda)R(N_2)-\lambda R(N_1)$, below $1.8$; at $\Jc$, where $N=80$ and $96$ exist, the line goes through $N=64$ and $96$ for $l\le5$ and through
$N=64$ and $80$ for $l=6$. The size error is the difference to the quadratic in $1/N^2$ through $N_2$, $N_1$ and the next smaller ring. The ratio of the
slopes between consecutive pairs of rings, which is $1$ for a pure $1/N^2$ law, has the median $1.04$ ($x=1/4$; interquartile range $1.00$--$1.10$),
$1.08$ ($x=1/9$; $0.98$--$1.24$), $1.27$ ($x=1/16$; $1.20$--$1.64$) and $1.02$ ($x=4/9$); the larger deviations at $x=1/16$ are captured by the size error.

\paragraph*{Bond dimension.}
The bond-dimension error is the change of the extrapolated value when the largest $\chi$ of both rings is replaced by the next smaller one (the
last step). Where the next smaller $\chi$ is far away (ratio below $0.74$, e.g.\ $256$ against $384$ for $l\le4$ at $0.10\le\Jt\le0.18$), the
change is multiplied by the ratio $\rho$ of the last step to the total change, measured on geometries with three bond dimensions: $\rho=0.37$
overall and $0.34$--$0.49$ for the individual block lengths. A term that cannot be formed is set to the median ratio of that term to the extrapolation
step in geometries of the same coupling and the same $d/l$. The measured $\rho$ corresponds to a ratio of about $0.6$ between consecutive changes;
for small changes ($\lesssim3\times10^{-5}$, mainly $l\le3$) the ratio is dominated by the DMRG convergence noise discussed below. If the
convergence in $\chi$ is geometric with this ratio, the residual beyond the largest $\chi$ is about $1.5$ times the last step. The bond-dimension term,
which dominates the error for $l\ge3$ (median $74$--$93\%$ of the squared error), may therefore be underestimated by up to this factor, and the total
errors by up to a factor $1.5$.

\paragraph*{Placement and convergence.}
The difference between the two placements of the blocks, shifted by one site, enters with the weights of the extrapolation, $(1+\lambda)$ and $\lambda$.
Of the $155$ ring calculations, $70$ ended at the maximal number of sweeps without meeting the stopping criterion (relative energy change $10^{-11}$ and
entropy change $10^{-8}$ between sweeps). Continuing two such rings ($N=64$, $\chi=256$, $\Jt=0.20$ and $0.28$) for $60$ further sweeps changed the energy
per site by at most $1.4\times10^{-11}$ and $R$ by at most $7.6\times10^{-6}$ and $9\times10^{-8}$, respectively. We add $u=7.6\times10^{-6}$ as the
uncertainty of every ring value, which contributes $(1+2\lambda)u$ to the extrapolated value. Geometries computed twice in independent runs agree to
$6\times10^{-12}$. The total error is the quadrature sum of the size, bond-dimension, placement and convergence terms. These are bounds rather than
standard deviations, and they are correlated between neighboring couplings and block lengths.

\paragraph*{Coefficients of the expansion in the coupling.}
For every ring size, bond dimension, geometry and placement that has all couplings of a window, $R$ (or the scalar share) is fitted by an unweighted
polynomial $\sum_jp_j\delta^j$ in $\delta=\Jc-\Jt$, of order $3$, $4$ or $5$, for the windows $0.20\le\Jt\le0.28$ and $0.22\le\Jt\le0.28$. With the
coupling written as $g_0(\delta)=g_c+s_1\delta+s_2\delta^2+\dots$, the coefficients of $R=\sum_kF_k(g_0-g_c)^k$ follow order by order from
$p_j=\sum_kM_{jk}F_k$, where $M_{jk}$ is the coefficient of $\delta^j$ in $(s_1\delta+s_2\delta^2+\dots)^k$; for example $F_1=p_1/s_1$ and
$F_2=(p_2-s_2F_1)/s_1^2$. The series of $g_0$ at $\Jc$ is obtained from a cubic through the five nearest points of the spectroscopy grid,
$s_1=1.7709$ and $s_2=4.95$ for the extrapolated coupling ($g_c=-3\times10^{-6}$) and $s_1=1.7787$, $s_2=4.92$ for the coupling of the largest ring
($g_c=0.0010$); a quartic through seven points changes $F_1$ by less than $10^{-5}$ and $F_2$ by less than $0.005$. The coefficients are then
extrapolated in $N$ with the rule above, with the convergence term multiplied by the norm of the corresponding row of the fit matrix (independent errors
at different couplings assumed). A variant is rejected if a weighted fit of the extrapolated values with their error bounds has probability below $1\%$;
the model error is the largest change among the variants that are not rejected. The central variant (order $4$, $0.20\le\Jt\le0.28$) is accepted for
every geometry.

\section{Explicit bond alternation}
\label{app:alternation}

\begin{figure}[t]
\centering
\figfile{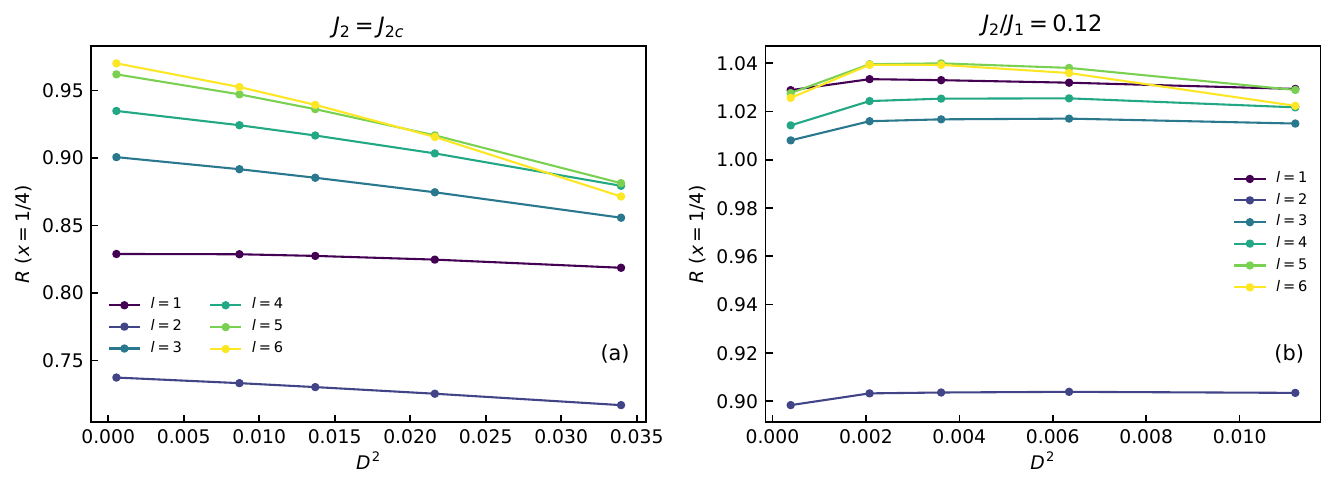}{\columnwidth}
\caption{Explicit bond alternation (earlier test, $\chi=128$, larger $\eta$): $R$ ($x=1/4$) versus the square of the induced dimer order, $D^2$, for $l=1$--$6$ at (a) $\Jc$ and (b) $\Jt=0.12$.}
\label{fig:alternation}
\end{figure}

At $\Jt=\Jc$ we add $\eta\sum_i(-1)^iJ_1\Sv_i\cdot\Sv_{i+1}$ with $\eta=0$, $5\times10^{-5}$, $10^{-4}$, $2.5\times10^{-4}$ and $5\times10^{-4}$ for
$\chi=128$, $160$ and $192$ (observable-converged iDMRG, blocks $l\le6$). Because the dimerization susceptibility diverges at $\Jc$, these small values
induce $D=0.055$--$0.117$, and $R(D)=R(0)+c\,D^2$ describes the data well for $D\le0.10$ [Fig.~\ref{fig:alternation}]. The slopes are similar for the three
bond dimensions, for example $c=-2.18$, $-2.31$, $-2.15$ ($x=1/4$, $l=6$) and $-5.50$, $-5.86$, $-5.46$ ($x=1/9$, $l=6$) for $\chi=128$, $160$, $192$, and the
extrapolations $R(0)$ agree with the undimerized $\chi=192$ state to $6\times10^{-4}$. Applied to the spontaneous dimerization of the production states,
however, these slopes overestimate its effect: the dimerized $\chi=160$ state ($D=0.028$) and the undimerized $\chi=192$ state differ in $R$ by
$4\times10^{-4}$ at $l=6$ ($x=1/4$), whereas $c\,D^2$ predicts $1.8\times10^{-3}$, and correcting each bond dimension with its own slope increases rather than
decreases the spread over $\chi$ ($1.4\times10^{-3}$ against $1.1\times10^{-3}$). An explicit alternation is a relevant perturbation that opens a gap; a
spontaneous finite-$\chi$ dimerization of nearly the same energy as the undimerized state is not, and the two need not affect the mutual information in
the same way. We therefore apply no alternation-based correction and base the quantitative analysis of the gapless side on periodic rings
(Sec.~\ref{sec:CFTresults}). At $\Jt=0.12$ (earlier test at $\chi=128$) the slopes are
consistent with zero.

\section{Measured marginal coupling}
\label{app:coupling}

Table~\ref{tab:coupling} lists the quantities from level spectroscopy that enter Sec.~\ref{sec:CFTresults}: the coupling $g(24)$ of Eq.~\eqref{eq:gN} on the
largest ring, the slope $b$ of a free fit of $1/g$ versus $\ln N$ ($N=12$--$24$) for $\Jt\le0.22$, and the bare coupling with the plateau slope $b=1.026$,
$g_0=g(N)/[1-b\,g(N)\ln N]$, either from the largest ring alone or extrapolated linearly in $1/N^2$ over $N=12$--$24$. The extrapolation changes $g_0$ by
$0.7\%$ at $\Jt=0.10$ and by up to $6\%$ close to $\Jc$; its zero lies at $0.24117$, that of the largest-ring coupling at $0.24175$. The extrapolated
coupling is used throughout; the other one serves to estimate the normalization uncertainty of the coefficients $F_k$ (Appendix~\ref{app:rings}). With the
bare distance $\Jc-\Jt$ instead of $g_0$, the curvature of the relation between the two enters the second-order coefficient.

\begin{table}[t]
\caption{Marginal coupling from level spectroscopy (ED, periodic rings $N=12$--$24$). $g_0^{(24)}$: from the largest ring; $g_0^{\rm ext}$: extrapolated
in $1/N^2$.}
\label{tab:coupling}
\begin{ruledtabular}
\begin{tabular}{lcccc}
$\Jt$ & $g(24)$ & $b$ (free) & $g_0^{(24)}$ & $g_0^{\rm ext}$ \\
\hline
0.00 & $0.2509$ & 0.989 & $1.380$ & $1.399$ \\
0.06 & $0.2088$ & 1.026 & $0.654$ & $0.655$ \\
0.10 & $0.1759$ & 1.063 & $0.4129$ & $0.4117$ \\
0.12 & $0.1576$ & 1.092 & $0.3244$ & $0.3230$ \\
0.14 & $0.1377$ & 1.136 & $0.2501$ & $0.2486$ \\
0.16 & $0.1160$ & 1.215 & $0.1866$ & $0.1851$ \\
0.18 & $0.0921$ & 1.377 & $0.1317$ & $0.1303$ \\
0.20 & $0.0657$ & 1.814 & $0.0836$ & $0.0822$ \\
0.22 & $0.0362$ & 3.868 & $0.0410$ & $0.0398$ \\
0.23 & $0.0201$ & -- & $0.0215$ & $0.0204$ \\
0.24 & $0.0031$ & -- & $0.0031$ & $0.0021$ \\
0.25 & $-0.0150$ & -- & $-0.0143$ & $-0.0153$ \\
0.26 & $-0.0342$ & -- & $-0.0308$ & $-0.0317$ \\
0.27 & $-0.0547$ & -- & $-0.0464$ & $-0.0472$ \\
0.28 & $-0.0766$ & -- & $-0.0613$ & $-0.0619$ \\
0.30 & $-0.1247$ & -- & $-0.0886$ & $-0.0891$ \\
\end{tabular}
\end{ruledtabular}
\end{table}

\section{Rényi-2 mutual information}
\label{app:renyi}

For the Rényi-2 mutual information we use $I^{(2)}_{\rm CFT}=-\tfrac14\ln(1-x)$ [Eq.~\eqref{eq:Icft}]. At these block sizes $R^{(2)}=I^{(2)}/I^{(2)}_{\rm CFT}$
is dominated by even--odd oscillations: at $\Jc$ ($\chi=192$, $x=1/4$) $R^{(2)}=1.06$, $1.93$, $1.01$, $1.86$ for $l=3$--$6$, and for $x=4/9$ the Rényi-2
mutual information can even be negative ($R^{(2)}=-0.20$ at $l=2$). Large alternating corrections are expected for Rényi entropies with $n>1$ in gapless
spin chains~\cite{Calabrese2010parity,CardyCalabrese2010}; they are much weaker in the von Neumann quantity used in the main text. Because the
odd-$l$ values ($1.06$, $1.01$) approach $1$, the data do not contradict $\mathcal{F}_2\equiv1$ at the self-dual radius, but block sizes $l\le6$ do
not allow a quantitative test.

\section{Numerical parameters and checks}
\label{app:params}

\begin{table*}[t]
\caption{Summary of numerical parameters.}
\label{tab:params}
\begin{ruledtabular}
\begin{tabular}{p{0.24\textwidth}p{0.68\textwidth}}
Item & value \\
\hline
Code & TeNPy 1.1 (iDMRG, finite DMRG, two-site, U(1)); exact diagonalization in fixed-$S^z$ sectors \\
Unit cell & 2 sites \\
Dimerized phase & 81 values $0.38\le\Jt\le0.70$; $\chi=48$--$256$ \\
MG vicinity & 25 values $0.498\le\Jt\le0.510$; $\chi=64,128,192$ \\
Lifshitz point (infinite) & grid spacing $10^{-4}$ around the root; $\chi=64$, $128$ \\
Gapless side (iMPS) & 13 values $0\le\Jt\le0.30$; $\chi=96$--$192$; alternation at $\Jc$ for $\chi=128$, $160$, $192$ \\
Channel decomposition & 17 values $0\le\Jt\le0.70$, $\chi=128$, $l\le3$, $d\le16$; $0.32\le\Jt\le0.45$: $d\le60$ ($\chi\le256$), $d\le200$ ($\chi=128$); $\Jc$, $\Jt=0$: $l\le6$ \\
Level spectroscopy & rings $N=12$--$24$; 31 values $0\le\Jt\le0.30$ \\
Rings, block MI & DMRG $N=32$, $48$, $64$ at $\Jt=0$, $0.10$--$0.22$, $0.23$, $\Jc$, $0.25$--$0.28$; $N=80$, $96$ at $\Jc$; $\chi=192$--$448$ ($\le320$ for $l=6$); $l\le6$, $d=l/2$, $l$, $2l$, $3l$ and largest gap \\
Lifshitz, finite size & open $N=40$--$320$ ($\chi=64$); rings: ED $N\le24$, DMRG $N=32$--$64$ ($\chi=160$; spacing $10^{-4}$ for $N=48$, $64$), $N=48$ with $\chi=240$, $320$ \\
Singular-value cutoff & $10^{-12}$ ($10^{-14}$ near MG; $10^{-10}$--$10^{-14}$ tested) \\
Blocks & $l\le3$ (dimerized), $l\le6$ (gapless) \\
Convergence & MI probes stable to $10^{-5}$ ($10^{-4}$ after 1500 sweeps) \\
Truncation error & $\le1.4\times10^{-8}$ (dimerized phase) \\
SU(2) check & $\max|3\langle S^zS^z\rangle-\langle\Sv\cdot\Sv\rangle|\le1.3\times10^{-5}$ \\
\end{tabular}
\end{ruledtabular}
\end{table*}

Exact checks: the energy per site is $-3/8$ to $10^{-12}$ at the MG point and agrees with the Bethe-ansatz value $\tfrac14-\ln2$~\cite{Hulthen1938} to
$2\times10^{-5}$ for $\Jt=0$, $\chi=48$; the energy computed from correlators agrees with the variational energy to $10^{-8}$. Exact diagonalization
reproduces the ground-state energy of the $N=12$ Heisenberg ring ($-5.387390917445$) and $E/N=-3/8$ at the MG point; DMRG on a ring of $N=12$ reproduces
the ED correlator to $2\times10^{-10}$ for both the ground state and the dimer partner state.

\bibliography{refs}

\end{document}